\documentclass[aps,pre,twocolumn,superscriptaddress,superscriptreference]{revtex4-2}

\usepackage{amsmath,bbold,bm,amssymb,scalerel,mathtools}
\usepackage{graphicx}
\usepackage{color}
\usepackage{enumitem}
\usepackage{algorithm,algpseudocode}
\usepackage{multirow}
\usepackage{colortbl,booktabs}
\usepackage{placeins}
\usepackage[usenames,dvipsnames]{xcolor}

\algnewcommand{\parState}[1]{\State%
	\parbox[t]{\dimexpr0.9\linewidth-\algmargin}{\strut #1\strut}
}
\algnewcommand{\Input}[1]{%
	\State \textbf{Input:}
	\Statex \hspace*{\algorithmicindent}\parbox[t]{.8\linewidth}{\raggedright #1}
}
\algnewcommand{\Output}[1]{%
	\State \textbf{Output:}
	\Statex \hspace*{\algorithmicindent}\parbox[t]{.8\linewidth}{\raggedright #1}
}
\algnewcommand{\Initialize}[1]{%
	\State \textbf{Initialize:}
	\Statex \hspace*{\algorithmicindent}\parbox[t]{.8\linewidth}{\raggedright #1}
}

\usepackage[colorlinks, linkcolor=BrickRed, urlcolor=blue!50!black, citecolor=blue!50!black]{hyperref}

\newcommand*{\citen}{}
\DeclareRobustCommand*{\citen}[1]{%
	\begingroup
	\romannumeral-`\x 
	\setcitestyle{numbers}%
	\cite{#1}%
	\endgroup
}

\newcommand\equalhat{%
	\let\savearraystretch\arraystretch
	\renewcommand\arraystretch{0.3}
	\begin{array}{c}
		\stretchto{
			\scalerel*[\widthof{=}]{\wedge}
			{\rule{1ex}{3ex}}%
		}{0.5ex}\\ 
		=%
	\end{array}
	\let\arraystretch\savearraystretch
}

\newcommand\diff{\mathrm{d}}

\newcommand{\gj}[1]{\textcolor{black}{#1}}

\usepackage{tikz}
\usetikzlibrary{arrows.meta,shapes,chains}
\tikzset{%
	>={Latex[width=2mm,length=2mm]},
	base/.style = {rectangle, rounded corners, draw=black,
		minimum width=4cm, minimum height=1cm,
		text centered, font=\sffamily},
	activityStarts/.style = {base, fill=blue!30},
	test/.style={base, diamond, aspect=2, text width=5em, fill=green!30},
	startstop/.style = {base, fill=red!30},
	activityRuns/.style = {base, fill=green!30},
	process/.style = {base, minimum width=2.5cm, fill=orange!15,
		font=\ttfamily},
}

\usepackage[normalem]{ulem}

\newcommand\add[1]{\textcolor{blue}{#1}}
\newcommand\delete[1]{\textcolor{red}{\sout{#1}}}

\begin{document}

\title{Computer Simulations and Mode-Coupling Theory of Glass-Forming Confined Hard-Sphere Fluids }

\author{Gerhard Jung}
\email{jung.gerhard@umontpellier.fr, gerhard.jung.physics@gmail.com}
\affiliation{Institut f\"ur Theoretische Physik, Universit\"at Innsbruck, Technikerstra{\ss}e 21A, A-6020 Innsbruck, Austria}
\affiliation{Laboratoire Charles Coulomb (L2C), Université de Montpellier, CNRS, 34095 Montpellier, France.}

\author{Thomas Franosch}
\email{Thomas.Franosch@uibk.ac.at}
\affiliation{Institut f\"ur Theoretische Physik, Universit\"at Innsbruck, Technikerstra{\ss}e 21A, A-6020 Innsbruck, Austria}

\begin{abstract}

We present mode-coupling theory (MCT) results for densely packed hard-sphere fluids confined between two parallel walls and compare them quantitatively to computer simulations. The numerical solution of MCT is calculated for the first time using the full system of matrix-valued integro-differential equations. We investigate several dynamical properties of supercooled liquids including scattering functions, frequency-dependent susceptibilities and mean-square displacements. Close to the glass transition, we find quantitative agreement between the coherent scattering function predicted from theory and evaluated from simulations, which enables us to make quantitative statements on caging and relaxation dynamics of the confined hard{-}sphere fluid.
\end{abstract}

\maketitle

\section{Introduction}

Physical confinement fundamentally changes the properties of simple liquids, including its structure \cite{Noyola1991,Nemeth:PRE59:1999,Nygard2017}, mechanical properties \cite{Granick1991MotionsAR,Mittal:PRL100:2008} and phase behavior \cite{exp:Pieranski1983,theo:Schmidt1996,theo:Schmidt1997,Alba-Simionesco_2006,Mandal2014}. Studying the impact of physical confinement on atomic, molecular or colloidal fluids is therefore of fundamental interest for science and technology. Systems of interest are for example crowded motion in living cells, liquids in porous media or micro- and nanofluidics. 

Arguably the simplest model for confined liquids are hard spheres or colloids, confined between two parallel walls. Within the fields of statistical and soft matter physics this system has been intensively studied via experiments \cite{C1SM06502E,PhysRevE.83.030502,PhysRevLett.99.025702,PhysRevLett.112.218302,doi:10.1063/1.4905472,PhysRevLett.108.037802,doi:10.1063/1.4825176,PhysRevLett.116.167801,PhysRevLett.117.036101,PhysRevLett.116.098302,PhysRevX.6.011014,villada2022layering}, computer simulations \cite{Noyola1991,Scheidler_2000,Scheidler_2002,doi:10.1021/jp036593s,PhysRevE.65.021507,PhysRevLett.85.3221,Baschnagel_2005,PhysRevLett.96.177804,doi:10.1063/1.2795699,doi:10.1021/jp071369e,Mittal:PRL100:2008,doi:10.1063/1.4959942,PhysRevLett.100.106001,doi:10.1063/1.3651478,PhysRevLett.111.235901,doi:10.1063/1.1524191,PhysRevE.86.011504,C3SM52441H,Mandal2017a,Jung2022_extreme} and theory \cite{theo:Schmidt1996,theo:Schmidt1997,Lang2012,Mandal2014,PhysRevX.6.011014, Jung:2020,Jung2020_cryst, Jung2020_self}. Important findings are that the confinement leads to inhomogeneous density profiles and anisotropic particle packing \cite{Nygard2017}. It also hinders or accelerates the dynamics depending on the wall-particle potential \cite{PhysRevLett.111.235901} and the wall roughness \cite{Scheidler_2002,doi:10.1063/1.3651478}. Additionally, confinement leads to a reentrant crystallization transition \cite{theo:Schmidt1996}, \gj{and confined hard disks have been shown to feature a fragile to strong liquid crossover \cite{Yamchi2012_disks}. }

Of particular interest in the field of soft{-}matter physics is the impact of confinement on the structural relaxation of dense liquids. It is well known that close to the glass transition, such dense liquids exhibit a drastic slowing down of transport upon compression or cooling. 
Since confinement has a significant impact on the packing and density of liquids it is therefore expected that it also strongly 
affects the glass transition itself. In a series of papers 
 the effect of physical confinement on the glass transition of a confined hard-sphere fluid has been investigated using mode-coupling theory \cite{Lang2010,Lang2012,Mandal2014,Jung:2020,Jung2020_self}. 
In particular, an intriguing multiple-reentrant glass transition 
has been observed \cite{Lang2010,Mandal2014}, in which a liquid can be vitrified 
and melted multiple times just by systematic reduction of the wall separation.
The recently developed cMCT \cite{Lang2010,Lang2012,Lang_2013}, an extension of mode-coupling theory (MCT) \cite{Bengtzelius_1984,Gotze2009,JansenReview2018} to confined liquids,  accounts for the broken translational invariance and the existence of multiple relaxation channels. 
Providing a numerical solution for the full time-dependent dynamics for this first-principle theory enabled us to qualitatively explain the multiple-reentrant scenario \cite{Jung:2020,Jung2020_self}. We have also found that all dynamical quantities connected to the glass transition, including structural relaxation time, diffusion coefficients and stretching coefficients also exhibit a non-monotonous dependence on the wall separation.

The theoretical analysis described in the above series of papers was, however, based on a technical numerical approximation. While this `diagonal approximation' was required to obtain stable solutions of the cMCT equations, it prevents quantitative comparison between theory and simulations. In the present manuscript we therefore present a methodology to overcome this technical approximation and provide time-dependent solutions for the full matrix-valued integro-differential cMCT equations. The methodology is based on mixing different discretization schemes to stabilize the numerical integration. To validate the theoretical results we perform event-driven computer simulations of hard spheres in confinement close to the glass transition \cite{Alder1957a,Rapaport1980,Bannerman2011,Mandal2014}. We find quantitative agreement between the cMCT results and the simulations for several important dynamical properties, including stretching coefficients and nonergodicity parameters. The present manuscript therefore not only validates the results obtained in previous manuscripts using the diagonal approximation for cMCT, but also presents non-trivial results from theory and simulations on particle caging and relaxation dynamics.

 Our manuscript is organized as follows. In Chapter \ref{sec:mct} we recapitulate the important cMCT equations and present the different algorithms used to solve these equations numerically.  Afterwards, we introduce the simulation model in Chapter \ref{sec:simulation}. In Chapter \ref{sec:struc_relax} the results for the structural relaxation at the critical packing fraction obtained from both cMCT and \gj{event-driven molecular dynamics simulations} are presented. In the subsequent chapters these results are then quantitatively compared with each other focusing on the nonergodicity parameter (see Chapter \ref{sec:nonergodicity}) and the critical exponent (see Chapter \ref{sec:vonSchweidler}). We summarize and conclude in Chapter \ref{sec:conclusions}.
  
  \section{Mode-coupling theory}
  \label{sec:mct}
  
    \begin{figure}
  	\includegraphics[scale=0.23]{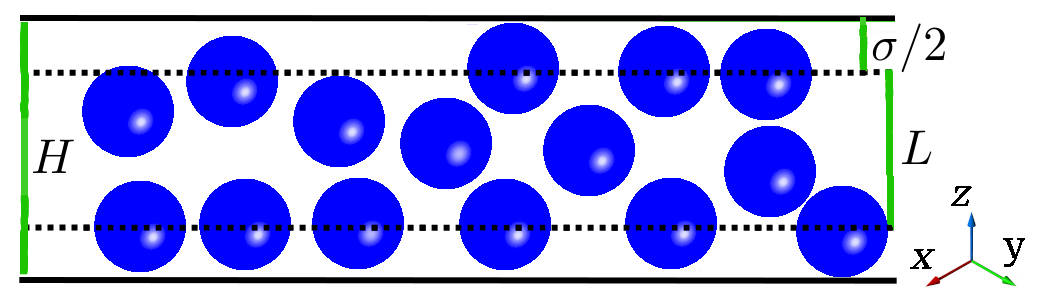}
  	\caption{Schematic of the hard{-}sphere fluid in confinement including the important length scales: the wall separation $ H $, the accessible slit width $ L = H - \sigma $ and the particle diameter $ {\sigma} $. The confinement direction is the $z$-axis. Adapted from Ref.~\cite{Jung2022_extreme}.  }
  	\label{fig:scetch}
  \end{figure}
  
  In this manuscript we will consider a system of hard spheres of diameter $\sigma$ confined between two parallel, hard and neutral walls at a distance $H$ (see Fig.~\ref{fig:scetch}). Due to the boundary regions, the particles can access only a reduced region in the slit of length $L=H-\sigma$, which will be the confinement length scale or accessible slit width reported in this manuscript. For this system, the packing fraction is defined as $ \varphi = N\pi \sigma^3/6V $, with particle diameter $ \sigma $, 
  volume $ V = A H $ and wall area $ A$.

  The original cMCT has been developed for Newtonian dynamics in Ref.~\cite{Lang2010} and extended to Brownian dynamics in Ref.~\cite{Schrack:2020a}. In the following chapter we will recapitulate the most important steps of the derivation and the integro-differential equations for Brownian dynamics. In Section \ref{sec:intro_time} we will then provide a detailed discussion of how to solve these equations numerically. Readers already familiar with these concepts or not interested in the technical details, may immediately jump to Chapter \ref{sec:simulation}.
  
  \subsection{cMCT equations of motion}
  
  To characterize the relaxation dynamics of the system we study the fluctuating density of $N$ particles in confinement (as sketched in Fig.~\ref{fig:scetch}),
  \begin{equation}
  \rho(\bm{r},z,t) := \sum_{n=1}^{N} \delta \left[\bm{r} - \bm{r}_n(t)\right] \delta\left[ z-z_n(t) \right] , 
  \end{equation}
  with the particle positions at time $t$ in lateral $ \bm{r}_n(t)=(x_n(t),y_n(t)) $ and transverse direction $ z_n(t) $. Our theoretical approach is then to derive the equations of motion for the incoherent scattering function,
  \begin{equation}\label{key}
  S_{\mu\nu}(q,t)  = \frac{1}{N} \left \langle  \rho_\mu(\bm{q},t)^* \rho_\nu(\bm{q},0) \right \rangle,
  \end{equation}
  which is defined from the fluctuating density modes,
  \begin{equation}\label{key}
  \rho_\mu(\bm{q},t) = \sum_{n=1}^{N} \exp\left[ {\rm i} Q_\mu z_n(t) \right] e^{ {\rm i} \bm{q}\cdot \bm{r}_n(t)}.
  \end{equation}
  Here we have introduced the wave vectors $ \bm{q}=(q_x,q_y) $ and discrete wavenumbers $ Q_\mu = 2 \pi \mu/L, \,\, \mu \in \mathbb{Z} $. In the following, the indices $ \mu  $  will be referred to as \emph{mode indices}. The modes naturally emerge due to the missing translational invariance in $z$-direction and the finite confinement length $L.$ The missing invariance also implies the emergence of an inhomogeneous density profile $ n(z) = \left \langle  \rho(\bm{r},z,t)  \right \rangle $ . We therefore also introduce the density Fourier amplitudes,
  \begin{equation}\label{key}
  n_\mu = \int_{-L/2}^{L/2} n(z) \exp \left[ {\rm i} Q_\mu z \right] \diff z,
  \end{equation}
  and  corresponding Fourier amplitudes $v_\mu$ for the local volume $ v(z) = 1/n(z) $.

  Based on the assumption of overdamped, colloidal dynamics without hydrodynamic interactions, the underlying microscopic equations of motion is described by the Smoluchowski equation \cite{Dhont1996,Schrack:2020a}. Using the Zwanzig-Mori projection operator formalism~\cite{Zwanzig2001,Gotze2009,Hansen:Theory_of_Simple_Liquids} and projecting on the density modes $\left\{ \rho_\mu(\bm{q},t) \right\} $ as set of distinguished variables, one finds,
\begin{align}\label{eq:eom1}
\dot{\mathbf{S}}(t)+\mathbf{D}\mathbf{S}^{-1}\mathbf{S}(t)
+\int_0^t \mathbf{\delta K}(t-t')\mathbf{S}^{-1}\mathbf{S}(t') \text{d}t' =0, 
\end{align}
with the diffusion matrix 
$\left[\mathbf{D}\right]_{\mu \nu} = D_0 (n^*_{\mu - \nu}/n_0) (q^2 + Q_{\mu} Q_{\nu})$, the short-time diffusion coefficient $ D_0 $ and the contracted force kernels $ \mathbf{\delta K}(t) $. Here and in the following, the explicit dependence on the wavenumber $ {q} $ is occasionally suppressed in the notation to improve readability.

Due to the decomposition of the density modes in the directions parallel and perpendicular to the walls, the relaxation also splits naturally into multiple channels \cite{Schrack:2020a}. We therefore introduce the contraction, 
\begin{align}\label{eq:contraction}
A_{\mu \nu}(q,t) &= \mathcal{C}\{\mathcal{A}_{\mu \nu}^{\alpha\beta}(q,t)\}\\
&:= \sum_{\alpha,\beta=\parallel,\perp}^{} b^\alpha(q,Q_\mu) \mathcal{A}_{\mu \nu}^{\alpha\beta}(q,t) b^\beta(q,Q_\nu), \nonumber
\end{align}
using the selector $ b^\alpha(x,z) = x \delta_{\alpha,\parallel} + z\delta_{\alpha,\perp}. $ This allows introducing the force kernels $ \delta\bm{ \mathcal{K}} $ as $ \delta  K_{\mu \nu}(q,t) = \mathcal{C}\{\delta \mathcal{K}_{\mu \nu}^{\alpha\beta}(q,t)\} $. The indices $ \alpha, \beta $ will be referred to as \emph{channel indices} and the matrix notation with calligraphic symbols $ \delta\bm{ \mathcal{K}} $ based on the super-index ($ \alpha,\mu $) will be employed. Having introduced the contraction, the equations of motion for the force kernel $\delta\bm{ \mathcal{K}} $ can be formally rewritten in terms of an irreducible memory kernel $\bm{ \mathcal{M}}$ \cite{Schrack:2020a},
\begin{align}\label{eq:eom2}
{\delta \bm{\mathcal{K}}}(t) = -\bm{\mathcal{D}} \bm{\mathcal{M}} (t) \bm{\mathcal{D}} - \int_0^t \bm{\mathcal{D}} \bm{\mathcal{M}}(t-t')   \delta \bm{\mathcal{K}}(t') \text{d}t' ,
\end{align}
Here, the channel diffusion matrices $ \left[ \bm{\mathcal{D}} \right]_{\mu \nu}^{\alpha \beta} = D_0 \delta_{\alpha \beta} n^{*}_{\mu - \nu}/n_0 $ has been introduced.

The equations of motion that have been derived so far are \emph{exact}. To make theoretical progress, however, a closure relation between the irreducible memory kernel $\bm{\mathcal{{M}}}(q,t)$ and the coherent scattering function $\mathbf{S}(q,t)$ has to be invoked. For this purpose an approximate mode-coupling functional \gj{$\mathcal{F}^{\alpha \beta}_{\mu \nu} \left[\mathbf{S}(t);q \right] = \mathcal{M}^{\alpha \beta}_{\mu \nu}(q,t)$} has been derived \cite{Schrack:2020a},
\begin{align}\label{eq:MCT_functional}
&\mathcal{F}^{\alpha \beta}_{\mu \nu} \left[\mathbf{S}(t);q \right] = \frac{1}{2N} \sum_{\substack{\bm{q}_1,\\\bm{q}_2=\bm{q}-\bm{q}_1}} \sum_{\substack{\mu_1,\mu_2\\\nu_1,\nu_2}} \mathcal{Y}^{\alpha}_{\mu \mu_1\mu_2}(\bm{q},\bm{q}_1,\bm{q}_2) \nonumber\\ &\times S_{\mu_1\nu_1}(q_1,t)S_{\mu_2\nu_2}(q_2,t)\mathcal{Y}^{\beta}_{\nu \nu_1\nu_2}(\bm{q},\bm{q}_1,\bm{q}_2)^*,
\end{align}
where the vertices $  \mathcal{Y}^{\alpha}_{\mu \mu_1\mu_2}(\bm{q},\bm{q}_1,\bm{q}_2) $ are smooth functions of the control parameters,
\begin{align}\label{eq:vertices}
&\mathcal{Y}^{\alpha}_{\mu \mu_1\mu_2}(\bm{q},\bm{q}_1,\bm{q}_2) = \frac{n_0^2}{L^4}\\
&\times \sum_\kappa v_{\mu-\kappa}^* \left[ b^\alpha(\bm{q}_1 \cdot \bm{q}/q,Q_{\kappa-\mu_2}) c_{\mu_1,\kappa-\mu_2}(q_1) + (1 \leftrightarrow 2)\right]. \nonumber
\end{align}

Conceptually the equations of motion for the incoherent intermediate scattering function $ \mathbf{S}^{(s)}(q,t) $ are identical to the ones for the coherent scattering function $ \mathbf(q,t) $ as introduced above. For details we refer to Refs.~\cite{Lang2014b,Schrack:2020a,Jung2020_self}.  The superscript $^{(s)}$ will refer in the following always to incoherent dynamics. In Ref.~\cite{Jung2020_self} we have also described how to calculate the in-plane mean-square displacement $ \delta r_\parallel^2(t) $ which we will also analyze in this work.

cMCT is purely based on three static input functions, the inhomogeneous density profile $ n(z) $ in transverse direction, the generalized structure factor $ S_{\mu\nu}(q)$, and the generalized direct correlation functions $ c_{\mu \nu}(q) $, which can all be determined numerically from first principle theories. The former is calculated via fundamental measure theory \cite{fmt:Rosenfeld1989,fmt:Roth2010} and the latter two via the generalized Ornstein-Zernike equation \cite{Hansen:Theory_of_Simple_Liquids} with Percus-Yevick closure. Details can be found in Refs.~\citen{Lang2010D,Petersen_2019}. Throughout this manuscript we use the same parameters as in Ref.~\cite{Jung:2020}.

\subsection{Nonergodicity parameter }
\label{sec:intro_nonergo}

The nonergodicity parameter is defined as the plateau value in the intermediate scattering function $ F_{\mu \nu} = \lim_{t\rightarrow \infty} S_{\mu \nu} $, which becomes non-zero above the mode-coupling glass{-}transition point. The nonergodicity parameter can be determined based the self-consistent equations \cite{Lang2010,Lang2012},
\begin{align}\label{eq:nonergo}
\bm{F}(q)=\left[ \bm{S}^{-1}(q)+\bm{S}^{-1}(q)\bm{K}[\bm{F}(q)]\bm{S}^{-1}(q)\right]^{-1} = \mathcal{I}[\bm{F}(q)],\\
\label{eq:nonergo2}\left[\bm{K}[\bm{F}(q)]\right]_{\mu \nu}= \sum_{\alpha,\beta=\parallel,\perp}b^\alpha(q,Q_\mu)\left[\bm{\mathcal{F}}^{-1}[\bm{F}(q);q]\right]_{\mu \nu}^{\alpha \beta}b^\beta(q,Q_\nu),
\end{align}
without solving the MCT equations of motion for the full time-dependence.    It has been proven that the above Eqs.~(\ref{eq:nonergo}) and (\ref{eq:nonergo2}) yield an iterative procedure, $ \bm{F}^{(n+1)}= \mathcal{I}\left[\bm{F}^{(n)}\right] $ which has a unique maximal solution $ \bm{\bar{F}}(q) $ starting from an initial condition $ \bm{F}^{(0)} = \bm{S}(q). $ This maximal solution $ \bm{\bar{F}}(q) $ can be identified as the nonergodicity parameter.    The critical point of the glass transition is then determined by solving this iteration for various packing fractions and identifying the critical point with the discontinuous jump in the nonergodicity parameter \cite{Lang2010}. We will report in Sec.~\ref{sec:struc_relax} the confinement-dependent critical packing fraction $\varphi_c(L)$, and in Sec.~\ref{sec:nonergodicity} the nonergodicity parameters for the coherent and incoherent scattering, $\mathbf{F}(q)$ and $\mathbf{F}^{(s)}(q)$, at the critical point.

\subsection{Numerical time integration }
\label{sec:intro_time}

 The mode-coupling{-}theory equations for the structural relaxation of confined colloids have been presented in Eqs.~(\ref{eq:eom1}) and (\ref{eq:eom2}). In principle, these equations could be discretized in time and then numerically integrated. The numerical solution is, however, with reasonable computational effort only stable up to times $ t \lesssim 0.1 \tau $, where $\tau = \sigma^2/D_0$ is the time needed for a free particle to diffuse its own diameter.
 
 In previous publications we have therefore introduced an effective memory kernel, ${\mathbf{M}}(t)$, via the implicit definition, 
 \begin{equation}\label{eq:def_meff}
 \hat{\mathbf{K}}(z) = -\left[ {\rm i} \mathbf{{D}}^{-1} + \hat{\mathbf{M}}(z)  \right]^{-1}.
 \end{equation}
 \gj{Here,  we have introduced $\hat{\mathbf{K}}(z) = \delta \hat{\mathbf{K}}(z) + \textrm{i} \mathbf{{D}}(q) $ with the high-frequency limit $\hat{\mathbf{K}}(z) \rightarrow \textrm{i} \mathbf{{D}}(q)$ as $z \rightarrow \infty$ as in Ref.~\cite{Schrack:2020a}.}
 Using this memory kernel, the final equation of motion for the structure factor can be rewritten as, 
 \begin{align}\label{eq:eomS}
 \mathbf{{D}}^{-1}\dot{{\mathbf{S}}}(t) +  \mathbf{S}^{-1} {\mathbf{S}}(t) + \int_{0}^{t} \mathbf{M}(t-t') \dot{\mathbf{S}}(t') \text{d}t' = 0,
 \end{align}
  which could be discretized and integrated accurately for arbitrarily long time scales (up to $t \lesssim 10^{20} \tau$) in Refs.~\cite{Jung:2020,Jung2020_self} using the diagonal approximation.
 
 \subsubsection{Diagonal approximation (DA) }
 
 The idea of the diagonal approximation is to simplify the equations of motion by neglecting all \gj{off-diagonal} terms of all matrices, $\mathcal{M}^{\alpha \beta}_{\mu \nu}(q,t) = \mathcal{M}^{\alpha}_{\mu }(q,t) \delta_{\alpha \beta} \delta_{\mu \nu}$, and ${S}_{\mu \nu}(q,t) = {S}_{\mu }(q,t) \delta_{\mu \nu}$. This not only significantly reduces the complexity of the equations and avoids large matrix multiplications, but it also enables the determination of the effective memory kernel directly from the mode-coupling functional in the time domain from Eq.~(\ref{eq:def_meff}) \cite{Schrack:2020a,Jung2020_self}. Using this DA we were able to make theoretical predictions for the structural relaxation of hard-sphere fluids in confinement which qualitatively agree with simulation results \cite{Mandal2014,Jung:2020}. 
 
 The DA has the character of a technical approximation and cannot be systematically lifted to investigate convergence or error estimates. It is therefore highly desirable to also solve the full matrix-valued system of equations, even it is only possible on smaller time scales, to validate the results obtained using the diagonal approximation.
 
 \subsubsection{Full solution (F)}
 
  The problem for the solution of the full matrix-valued mode-coupling{-}theory equations is that the definition Eq.~(\ref{eq:def_meff}) combined with Eq.~(\ref{eq:eom2}) cannot be converted into a closed differential equation in the temporal domain for the effective memory kernel. The reason for this is that the matrix multiplication is not commutative.
  
   {An} alternative route (in the following called route A) would then be to evaluate $ {\delta \bm{\mathcal{K}}}(t)  $ via Eq.~(\ref{eq:eom2}), use this solution to determine the effective memory kernel via the definition Eq.~(\ref{eq:def_meff}) in time-domain and finally integrate Eq.~(\ref{eq:eomS}). This improves the stability but still restricts the stable solution to a time scale $ t \lesssim 10 \tau, $ which is not sufficient to study the structural{-}relaxation process close to the glass transition. 
  
  Here, we propose a novel algorithm which enables stable solutions up to times $ t \approx 10^5 \tau. $ The fundamental idea for this algorithm is based on the observation described in Ref.~\cite{Jung2020_self}, Appendix A3, that the numerical solution of the velocity{-}autocorrelation function (VACF) is more stable if it is determined as the second time-derivative of the mean{-}square displacement than it is if being integrated directly. Since the VACF is equivalent to the $ \mu=\nu=0 $ mode of the current kernel, $ Z(t) =  {\delta {\mathcal{K}^{\parallel\parallel}_{00}}}(t) $, it is a natural idea to introduce a generalized mean-square displacement $ \delta \mathcal{\ddot{R}}^{\alpha \beta}_{\mu \nu}(t) = {\delta {\mathcal{K}^{\alpha \beta}_{\mu \nu}}}(t) $ for which we can derive the equation of motion,
  \begin{equation}\label{eq:msd_general}
  \delta \bm{\mathcal{{R}}}(t) + \bm{\mathcal{D}}\int_0^t  \bm{\mathcal{M}}(t-t') \delta \bm{\mathcal{{R}}}(t') \text{d}t' = \bm{\mathcal{D}} t.
  \end{equation} 
  
   The proposed ``route B'' is to replace the evaluation of Eq.~(\ref{eq:eom2}) in route A with the integration of Eq.~(\ref{eq:msd_general}), followed by the determination of the effective memory kernel using the generalized mean-square displacement. Details on the integration and discretization scheme of the coupled integro-differential equations introduced above are described in Appendix~\ref{app:integration}. The numerical evaluation of the MCT equations also requires {a} discretization of the wavenumbers $q$ and the introduction of a mode cutoff $|\mu| \leq M$. Technical details on these numerical requirements are given in Appendix \ref{sec:intro_numerics}.
   
  The MCT results reported in Sec.~\ref{sec:struc_relax} will be extracted at packing fractions just below the critical point, which has been determined in Sec.~\ref{sec:intro_nonergo}. We will discuss the coherent scattering function $\mathbf{S}^{c}(q,t)$, the incoherent scattering function  $\mathbf{S}^{s}(q,t)$ as well as the mean-square displacement $\delta \mathcal{R}_{00}^{\parallel \parallel}(t).$

\section{ Simulation model}
  \label{sec:simulation}
  
  We perform event-driven \gj{molecular dynamics (EDMD)} simulations of hard spheres confined between parallel, hard and neutral walls in the \emph{NVT} ensemble \cite{edmd:alder1957,edmd:RAPAPORT1980,edmd:bannerman2011}. We choose five different confinement lengths $ \gj{L/ \sigma=1.0,1.25,1.5,1.75,2.0 }$ and as packing fraction the critical packing fraction $\varphi_c(L)$ determined in Ref.~\cite{Mandal2014}. The total number of particles varies slightly with $L$ but is roughly $N=5000$. To prevent crystallization, we introduce polydispersity and draw radii from a Gaussian distribution of mean $\sigma$ and variance $0.15 \sigma$. It should be noted that the polydispersity leads to additional small deviations between simulations and cMCT (monodisperse), which have been discussed in Ref.~\cite{Mandal2014}. 
  
  Using event-driven simulations implies that the microscopic dynamics is Newtonian. This stands in contrast to the microscopic Brownian dynamics used for cMCT. We made this choice because cMCT is numerically less stable for inertial dynamics in the density fluctuations and including Brownian motion to event-driven systems is very peculiar. The above discrepancy is, however, not problematic since we are only interested in the slow, structural relaxation and the results are thus expected to be independent of the microscopic dynamics \gj{(as has been shown in 3D bulk systems \cite{PhysRevLett.81.4404}, and in confinement using cMCT with DA \cite{Jung2020_self})}.

  \section{Structural relaxation in confined supercooled liquids}
  \label{sec:struc_relax}
  
     \begin{figure}
  	\includegraphics[scale=1]{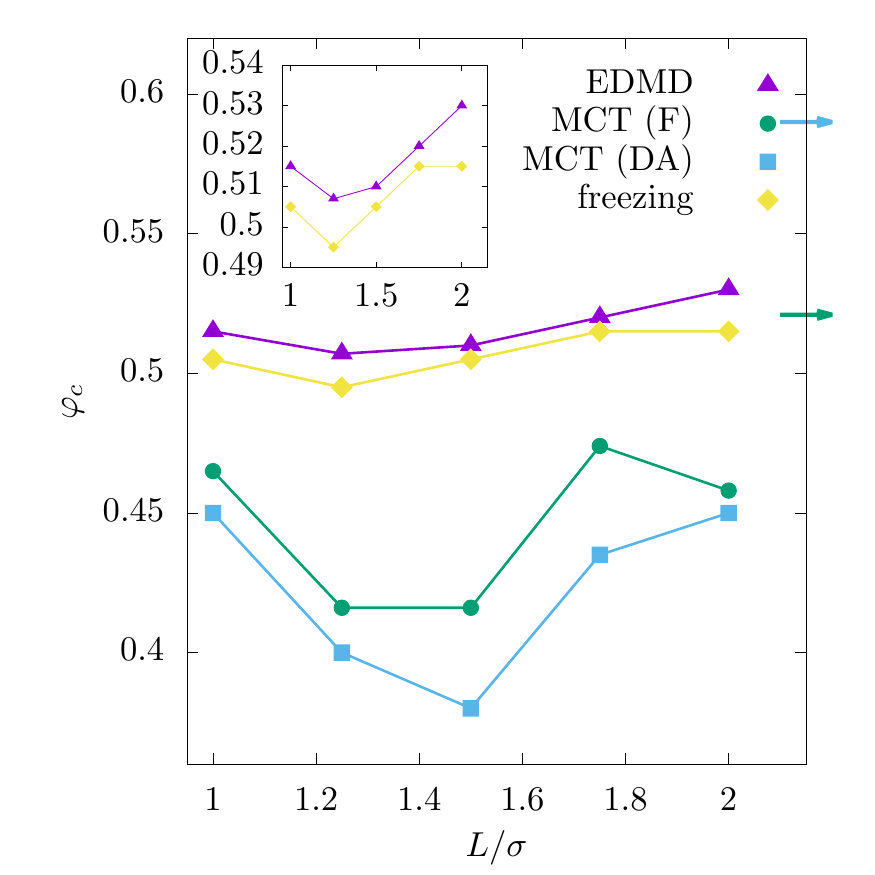}
  	\caption{Critical packing fractions $ \varphi_c $ of the glass transition for different accessible slit widths $ L $ as determined from theory (MCT) and computer simulations (EDMD, \cite{Mandal2014}). Results are shown for MCT using the full model (F) and the diagonal approximation (DA, \cite{Jung:2020}) as well as for the freezing transition as calculated in Ref.~\cite{Jung2020_cryst}. The arrows indicate the results in the bulk limit and the inset visualizes a zoomed section of the plot. }
  	\label{fig:critical}
  \end{figure}
  
  Mode-coupling theory predicts a transition from an ergodic liquid phase, in which the coherent scattering functions decays to 0 for $t\rightarrow \infty$, to a nonergodic glass phase upon compression \cite{Bengtzelius_1984,Gotze2009}. We report the critical packing fraction $\varphi_c$ predicted by cMCT in Fig.~\ref{fig:critical}. As has been reported in Refs.~\cite{Lang2010,Mandal2014}, cMCT predicts a reentrant glass transition, in which $\varphi_c$ depends non-monotonically on $L.$ Here, we show that the same scenario is found when using the full cMCT without diagonal approximation which predicts a slightly larger critical value, likely due to the inclusion of additional relaxation channels that were neglected within DA.
  
  As is commonly known, hard-sphere glasses studied by computer simulations avoid the ergodicity-breaking transition predicted by MCT \cite{Kob1994}. Nevertheless, a drastic slow down of transport can be observed which can be described over several orders of magnitude by a power-law divergence. In this way it is possible to extract a ``critical'' packing fraction $\varphi_c$ for the glass transition, as has been done in Ref.~\cite{Mandal2014}. Comparing simulations and cMCT it can be observed that cMCT significantly underestimates the critical packing fraction, but that both show qualitatively the same reentrant scenario (see Fig.~\ref{fig:critical}). This reentrant scenario is mainly induced by structural rearrangements. At confinement lengths corresponding to integer multiples of the particle diameter, $L \approx n \sigma,$ $n\in \mathbb{N}$, particle are packed into $n$ pronounced layers (commensurate packing) which allows for relatively unhindered in-plane diffusion and thus leads to larger critical packing fractions. For intermediate $L$ particles between the layers hinder diffusion (incommensurate packing), thus increasing the coupling between the layers and restricting the particle motion \cite{villada2022layering}. The logical consequence are smaller critical packing fractions $\varphi_c$ \cite{Mandal2014}. The amplitude of the oscillations visible in the critical packing fraction is significantly stronger for cMCT compared to EDMD. This is likely due to the difference in dispersity (monodisperse vs. polydisperse) which correspondingly reduces the impact of the non-monotonous particle packing described above \gj{\cite{Mandal2014}}.

  In the following, we will perform cMCT calculations and event-driven simulations for the five different values of $L$ slightly below their respective critical point $\varphi(L) = \varphi_c(L) - 0.005.$ Despite the introduction of polydispersity it has been shown in Ref.~\cite{Jung2020_cryst} using an enhanced equilibration SWAP Monte Carlo algorithm that confined systems are prone to crystallization. Here, we therefore intentionally do not use SWAP to equilibrate the simulations implying that the systems are slightly supercooled. \gj{We find some early signs of crystallization such as the emergence of small peaks in the structure factor $S_{00}(q)$ visible in Ref.~\cite{Mandal2017a} Fig.~1(c)}. However, the effect should be small enough to ensure only a marginal impact on the relaxation dynamics analyzed below \cite{Jung2020_cryst, Mandal2017a}.
  
     \begin{figure}
  	\includegraphics[scale=1]{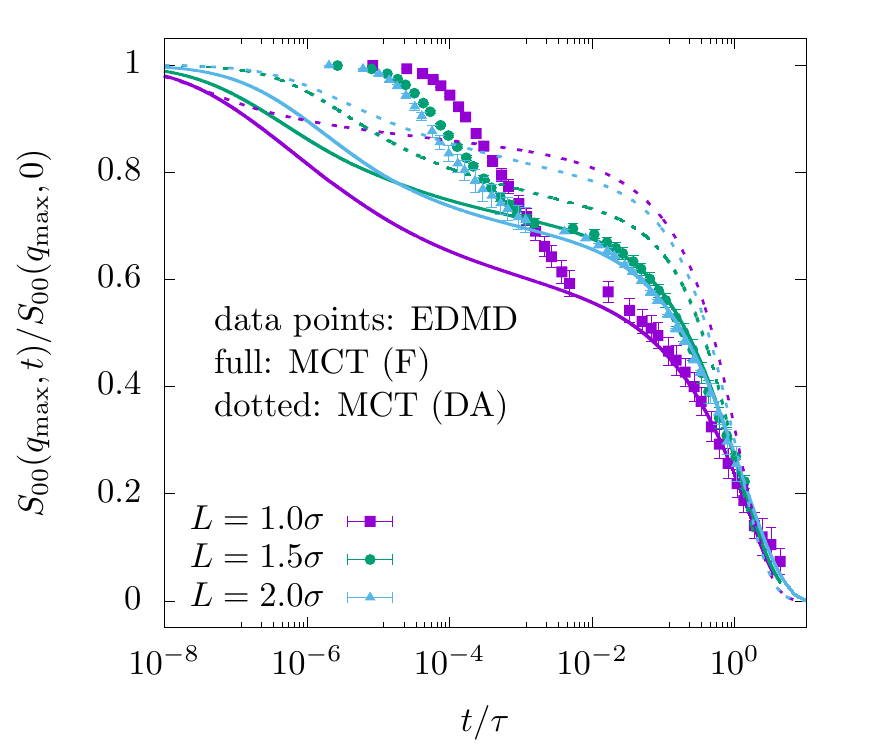}
  	\caption{Normalized coherent intermediate scattering function $ S_{00}(q_\text{max},t) $ for different accessible wall widths $ L $ for the wavenumber $ q_\text{max}\sigma \approx 2 \pi   $ which corresponds to the first peak in the structure factor. The structural relaxation is fitted with a KWW law, Eq.~(\ref{eq:KWW}), to extract the structural relaxation time $ \tau $ (plotted in thin lines). }
  	\label{fig:coherent_scattering}
  \end{figure}
  
The coherent scattering function, characterizing the decay of density correlations slightly below $\varphi_c$ shows the typical two-step relaxation known for supercooled liquids \cite{VanMegen1993,Kob1994,Gotze2009} (see Fig.~\ref{fig:coherent_scattering}); a short-time relaxation due to thermal motion is followed by a plateau which decays on much longer time scales. \gj{The short-time relaxation depends on the microscopic dynamics and is thus different for EDMD and cMCT.} The height of the plateau then contains information on the length scale and strength of the cages in which the particles are trapped before structural relaxation sets in. It can be identified with the nonergodicity parameters introduced above in Sec.~\ref{sec:intro_nonergo}. Here, we observe that the nonergodicity parameters change non-monotonically with the confinement length scale $L$. Importantly, we find nearly quantitative agreement for the plateau height and the long-time structural relaxation between the full cMCT solution and computer simulations. This is a rather non-trivial result for a first-principle theory without fitting parameters and the most important result of this paper. Figure~\ref{fig:coherent_scattering} also highlights that the diagonal approximation is not able to predict the correct values for the nonergodicity parameters.

   \begin{figure}
	\includegraphics[scale=1]{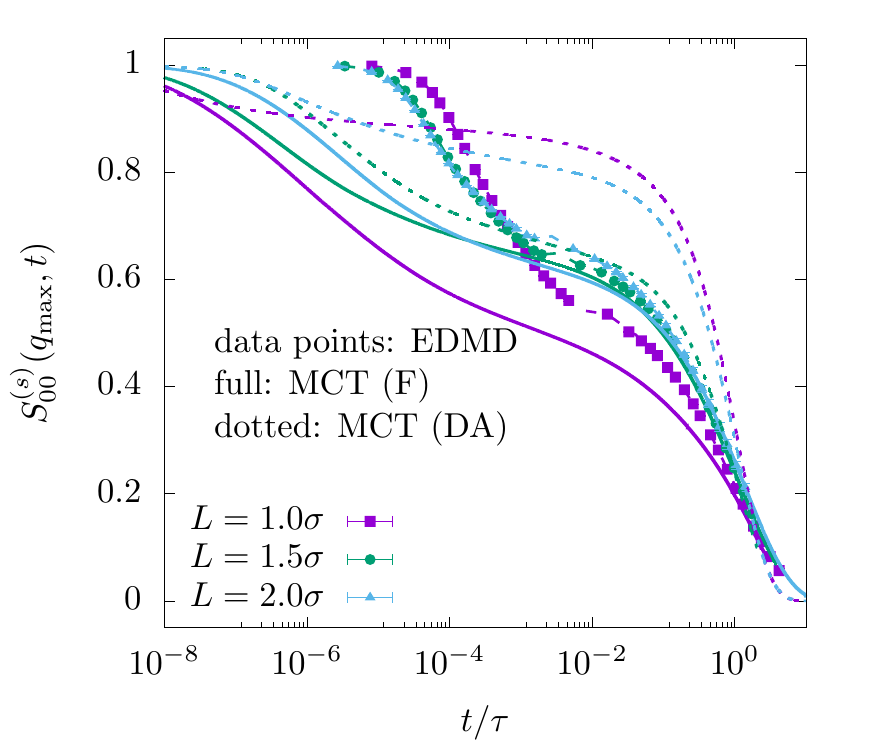}
	\caption{Incoherent intermediate scattering function $ S^{(s)}_{00}(q_\text{max},t) $ for different accessible wall widths $ L $ for the wavenumber $ q_\text{max}\sigma \approx 2 \pi   $. The structural relaxation is fitted with a KWW law, Eq.~(\ref{eq:KWW}), to extract the structural relaxation time $ \tau $ (plotted in thin lines).  }
	\label{fig:incoherent_scattering}
\end{figure}

To further characterize structural relaxation, we also calculate the incoherent intermediate scattering function describing the decay of the single-particle density,
\begin{equation}\label{key}
S^{(s)}_{\mu\nu}(q,t)  = \frac{1}{N} \left \langle \sum_{n=1}^N  \exp\left[ {\rm i} Q_\mu (z_n(t) - z_n(0) \right] e^{ {\rm i} \bm{q}\cdot (\bm{r}_n(t) - \bm{r}_n(0)} \right \rangle.
\end{equation}
Importantly, it can be observed that the incoherent scattering function features the same two-step decay discussed for $S_{00}(q,t)$ only with different values of the nonergodicity parameters (see Fig.~\ref{fig:incoherent_scattering}). The same statement actually holds for all higher-order modes, $S_{\mu \nu}(q,t)$ and $S^{(s)}_{\mu \nu}(q,t)$. This finding is consistent with previous results for structural relaxation in bulk \cite{Fuchs1998} and in confinement \cite{Jung2020_self} and consistent with our scaling analysis of mode-coupling theories with multiple relaxation channels close to criticality in Ref.~\cite{Jung:2020_scalingtheory}. The agreement between simulations and cMCT is not as good as for the coherent intermediate scattering function, in particular, for $L=1.0\sigma.$ However, for the other confinement lengths $L$ the result is significantly better. Also in this case the full cMCT solution improves the prediction from the DA.

  To extract the nonergodicity parameters, all incoherent and coherent scattering functions calculated for different confinement lengths $L$ and wavenumbers $q$ are fitted by the phenomenological Kohlrauch-William-Watts (KWW) law \cite{phillips1996stretched}
\begin{equation}\label{eq:KWW}
S_{\mu \nu}(q,t) = F_{\mu \nu}(q) \exp\left[ - (t/\tau)^{\beta_\text{kww}} \right],
\end{equation}
to extract the structural relaxation time $\tau(L),$ which is used to unify the time scales between the cMCT solutions and the simulations throughout the manuscript. It also enables us to determine important properties of the structural{-}relaxation process.  We will analyze the nonergodicity parameters $F_{\mu \nu}(q)$, $F^{(s)}_{\mu \nu}(q)$ and stretching coefficients $\beta_\text{kww}$ in Secs.~\ref{sec:nonergodicity} and \ref{sec:vonSchweidler}, respectively.

   \begin{figure}
	\includegraphics[scale=1]{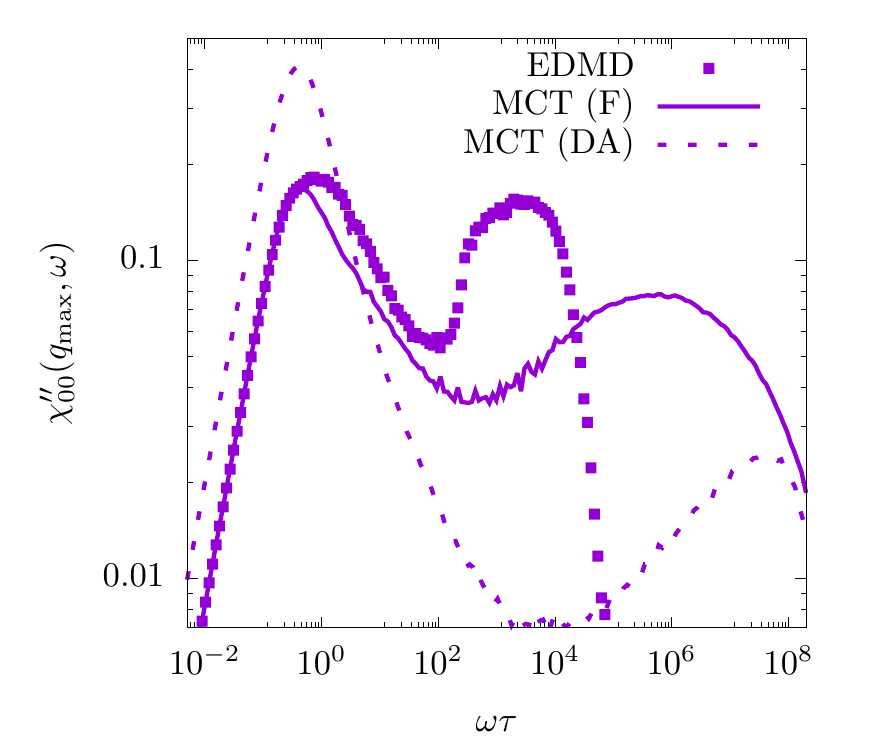}
	\caption{ Frequency-dependent susceptibility $ \chi_{00}^{\prime \prime}(q_\text{max},\omega)$ calculated from the in-plane ISF, Eq.~(\ref{eq:susceptibility}),  for $ L=1.0\sigma $ at wavenumber $ q_\text{max} \sigma \approx 2 \pi  $. The data has been extracted from the incoherent ISF due to improved statistics for the EDMD simulations.   }
	\label{fig:susceptibility}
\end{figure}  

The two-step relaxation scenario is also known from scattering experiments extracting frequency-dependent susceptibilities of glassy materials \cite{VanMegen1993,Franosch1997}. Based on the calculated scattering functions, we can identify the susceptibility,
\begin{equation}\label{eq:susceptibility}
\chi^{\prime \prime}_{00}(q,\omega) = \omega \hat{S}_{00}(q,\omega),
\end{equation}  
using the Fourier cosine transform, $ \hat{S}_{00}(\omega) = \int_0^\infty \text{d} t \cos(\omega t) S_{00}^{(s)}(q,t)  $ \cite{Franosch1997}. The results clearly show two distinct peaks in the susceptibility, which correspond to the two-relaxation processes (see Fig.~\ref{fig:susceptibility}). From the high-frequency flank of the low-frequency peak it is possible to extract the exponent $b$ of the von Schweidler law, also know as the late $\beta$ relaxation \cite{Gotze2009},
\begin{equation}\label{eq:schweidler}
S_{\mu \nu}(q,t) = F_{\mu \nu}(q) - A(t/\tau)^b.
\end{equation}
 The von Schweidler law has been known from scattering experiments \cite{doi:10.1063/1.4770055} and can be predicted using the celebrated $ \beta $-scaling equation, which can be derived from first principles using MCT \cite{Gotze_1990,PhysRevE.52.4134}. For the actual fitting in this work we use the late $ \beta $-relaxation to higher order \cite{Franosch1997},
\begin{equation}\label{eq:schweidler2}
S_{\mu \nu}(q,t) = F_{\mu \nu}(q) - A(t/\tau)^b + B (t/\tau)^{2b} {.}
\end{equation}
We find that the von Schweidler exponent $b$ coincides with the stretching exponent $\beta_\text{kww}$ at wavenumber $ q_\text{max}\sigma \approx 2 \pi   $ within the accuracy of the fitting procedure \cite{PhysRevA.45.898,FUCHS1994241,doi:10.1080/00411459508203937}.

In previous work based on the DA we have found an interesting kink on the right flank of the low-frequency peak \cite{Jung:2020}, which can be explained by the multiple relaxation channels in the lateral and the transverse direction. Having performed additional computer simulations and after solving the full cMCT equations we can conclude that this observation {may} be a numerical artifact {of the DA}. It is likely that the mode mixing introduced by including the off-diagonal components leads to additional couplings between the relaxation channels and thus removes the multi-step relaxation induced by the multiple relaxation channels.

   \begin{figure}
	\includegraphics[scale=1]{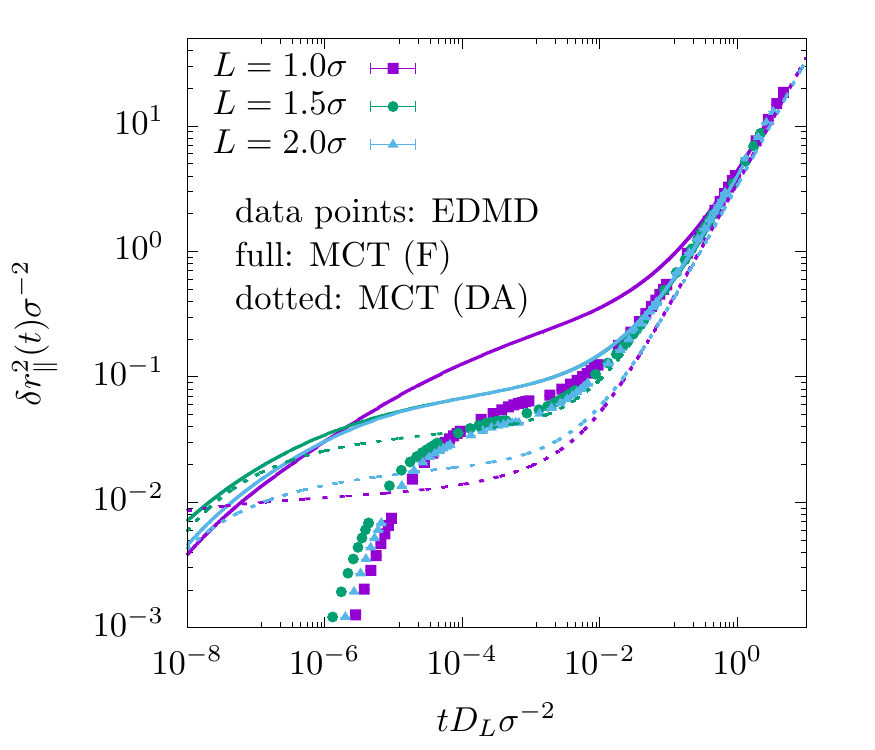}
	\caption{In-plane mean-square displacement $ \delta r_\parallel^2(t) $ for different accessible wall widths $ L $. The long time diffusion coefficient $ D_L $ is fitted to the curves to adjust the time scales (\ref{eq:schweidler_msd}).  }
	\label{fig:msd}
\end{figure}

Another {fundamental} dynamical observable to characterize structural relaxation is the mean{-}square displacement (MSD), $\delta r_\parallel^2(t) = \langle \sum_{n=1}^N (\bm{r}_n(t) - \bm{r}_n(0))^2  \rangle. $ here taken only for the 2D lateral motion. It is easily accessible in simulations and the extracted long-time diffusion coefficient {$D_L := (1/4) \lim_{t\to\infty}\diff \delta r_\parallel^2 /\diff t $} is a popular measure to characterize the glass transition \cite{Kob1994,Gotze2009,Mandal2014}. Unsurprisingly, the MSD also shows a two-step behavior. At short times, the motion is inertial for the EDMD simulations (and diffusive for cMCT), followed by a slower increase at microscopic length scales, characterized by a `cage size' or localization length $l_\parallel$. At long times, the motion becomes diffusive again (see Fig.~\ref{fig:msd}). This behavior is qualitatively the same for simulations and theory. Combining cMCT with the DA one, however, observes a systematic underestimation of the localization length. Therefore, we extract the important properties of the relaxation dynamics using {an interpolating} equation to fit the MSD on intermediate and long times,
\begin{equation}\label{eq:schweidler_msd}
\delta r_\parallel^2(t) = 4 l_\parallel^2 - A(t/\tau)^b + B (t/\tau)^{2b} + 4 D_L t.
\end{equation}
Here the factors of 4 account for the two-dimensional lateral motion. The equation thus connects the late $ \beta $-relaxation \cite{Franosch1997} with the long-time diffusive regime. We will analyze {$l_\parallel = l_\parallel(L)$ as a function of the confinement length $L$} in the following section.

  \section{ Nonergodicity parameters }
  \label{sec:nonergodicity}
  
  In the previous section we have discussed the time-dependent structural relaxation {displaying} a very similar two-step relaxation behavior with a characteristic plateau across different confinement lengths $L$, wavenumbers $q$ and mode indices $\mu.$  In the following, we will analyze and discuss the nonergodicity parameter (NEP) and localization length {characterizing} the plateau height.
  
    \begin{figure}
  	\includegraphics[]{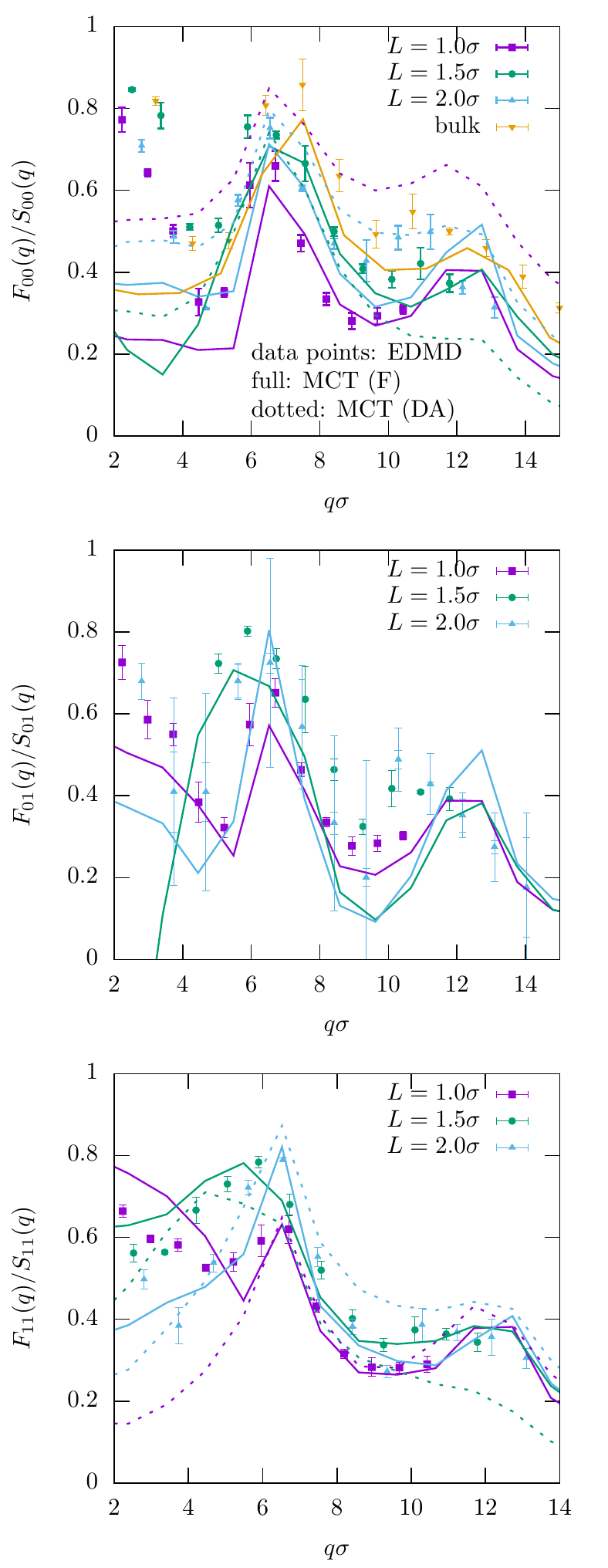}
  	\caption{Normalized critical nonergodicity parameter $ F_{\mu \nu}^\text{crit}(q)/S_{\mu \nu}(q) $ for different accessible slit widths $ L $ for the in-plane mode $ \mu=0,\nu=0 $ (top panel), the first off-diagonal mode $ \mu=0,\nu=1 $ (middle panel) and the first diagonal mode $ \mu=1,\nu=1 $ (bottom panel). Results are shown for MCT using the full model (F) and the diagonal approximation (DA) and for even{t}-driven molecular dynamics simulations (EDMD) using a KWW fit of the coherent intermediate scattering function $S_{\mu \nu}(q,t)$. }
  	\label{fig:nonergo_coh}
  \end{figure}
  
  The normalized NEP $F_{\mu \nu}(q)/S_{\mu \nu}(q)$ shows a very similar $q-$dependence as the structure factor, with a pronounced peak at roughly $q \sigma \approx 2\pi$ (see Fig.~\ref{fig:nonergo_coh}). It should be emphasized that this is a non-trivial result, since the normalization removes any direct impact of the structure factor.  In fact, a similar NEP is known from bulk liquids (see orange \gj{(light gray) line and downwards-pointing triangles} in Fig.~\ref{fig:nonergo_coh}, top panel). The peaks in the NEP highlight the importance of the cages, formed by nearest neighbors, for the dynamical particle trapping. We {conclude}, that this observation remains valid in confinement, which only quantitatively modifies the NEPs. Consistent with Ref.~\cite{Mandal2017a} we find that the nonergodicity parameters vary non-monotonically with confinement length $L$ and that the diagonal approximation significantly overestimates the structural arrest. The latter has been accounted to the effect of polydispersity in Ref.~\cite{Mandal2017a}. Using the full solution of the cMCT equations we can clearly draw a different conclusion in the present manuscript and account this to the diagonal approximation itself. When using the matrix-valued equations we observe good quantitative agreement between simulations and cMCT over all  confinement lengths $L$ and wavenumbers. \gj{The discrepancy at small wavenumbers $q\sigma \lesssim 4$ can be explained by the differences in polydispersity, as shown for bulk fluids in Fig. 5 of Ref.~\cite{PhysRevE.82.011504}}. We further find that the NEPs in confinement are consistently smaller than in bulk fluids. We interpret these results by additional and anisotropic constraints due to the walls in the $z-$direction, which restrict the motion even stronger than what we observe here in the in-plane direction and thus induce the dynamical arrest.
  
  Analyzing the nonergodicity parameters in the off-diagonal, $F_{01}(q)$, and the higher-order modes, $F_{11}(q)$ we can characterize some of the anisotropic effects on the cages. We observe that also these terms are dominated by the peak at {$q \sigma \approx 2\pi$}, highlighting the importance of the particle cages. \gj{Interestingly, the peak for incommensurate packing $L=1.5\sigma$ is broader and slightly shifted to smaller $q$. We explain this with anisotropic cages and larger fluctuations of cages sizes at incommensurate packing. This effect could be induced by the reduced overall packing fraction while some particles are being locked between the two pronounced layers thus significantly slowing down the dynamics.} \gj{In $F_{11}(q)$ also a second peak at smaller $q \sigma <  4$ is emerging for $L=1.0\sigma$} indicating the effect of the boundary itself. These effects are quantitatively reproduced by both simulations and cMCT, showing that the first-principle theory is indeed able to predict the complex shape of the cages and the trapped motion of the particles in confinement.

  \begin{figure}
	\includegraphics[]{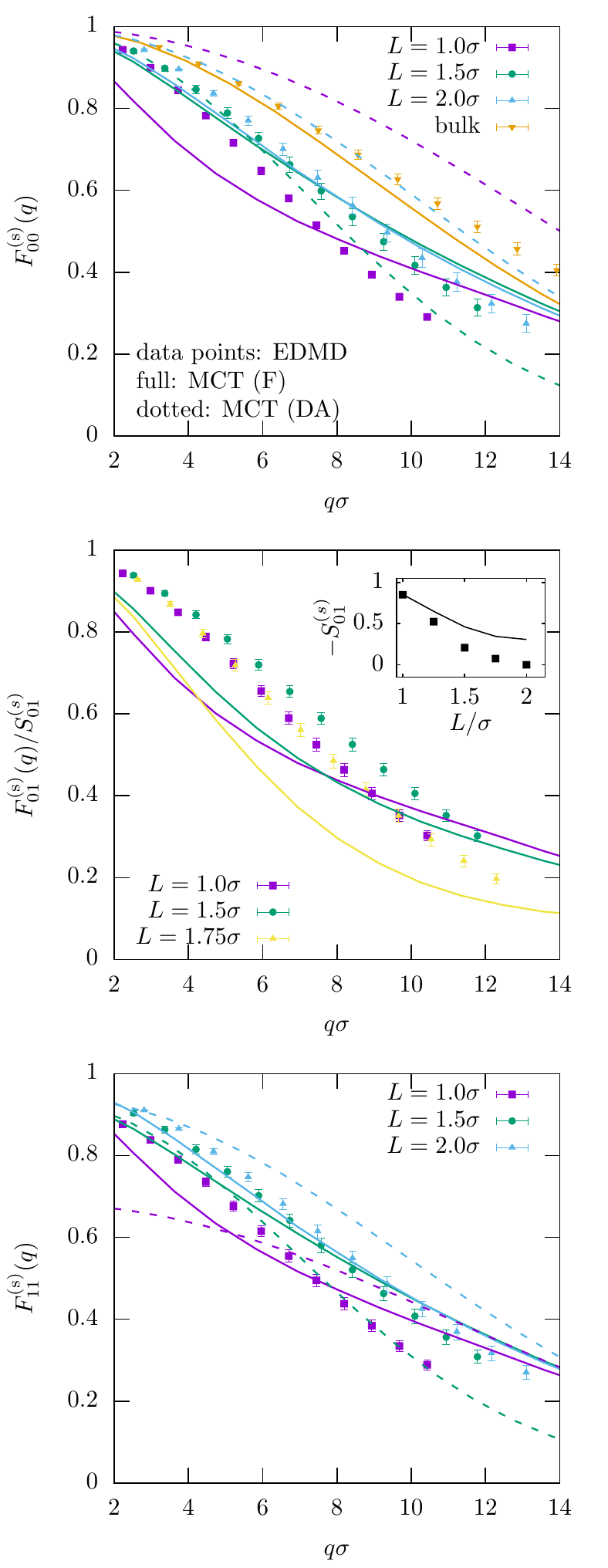}
	\caption{
Normalized critical incoherent nonergodicity parameter $ F^\text{(s)}_{\mu \nu}(q) $ for different accessible slit widths $ L $ for the in-plane mode $ \mu=0,\nu=0 $ (top panel), the first off-diagonal mode $ \mu=0,\nu=1 $ (middle panel) and the first diagonal mode $ \mu=1,\nu=1 $ (bottom panel). Results are shown for MCT using the full model (F) and the diagonal approximation (DA) and for even-driven molecular dynamics simulations (EDMD) using a KWW fit of the incoherent intermediate scattering function $S^\text{(s)}_{\mu \nu}(q,t)$.  Middle panel shows the normalized quantity normalization $S_{01}^{(s)}$ as inset and it holds generally $S_{00}^{(s)} = S_{11}^{(s)} = 1$ \cite{Lang2014b}.  }
	\label{fig:nonergo_incoh}
\end{figure}

Different from the normalized coherent nonergodicity parameter its incoherent counterpart, $F^{s}_{\mu \nu}(q)/S^{s}_{\mu \nu}$  decays monotonically with the wavenumber $q,$ see Fig.~\ref{fig:nonergo_incoh}. This reflects the localization and strong caging of the particle.  Consistent with {the} observations discussed above we find that the incoherent NEPs are smaller in confinement than they are in bulk. The agreement between cMCT and simulations is not as good as for the case of the coherent NEPs, but the important qualitative features are very well reproduced and only small quantitative deviations can be observed. Importantly, we find that the full matrix-valued solution outperforms the diagonal approximation. The off-diagonal component $F^{s}_{1 0}(q)$ and first mode $F^{s}_{1 1}(q)$ are nearly indistinguishable from the zero-mode $F^{s}_{0 0}(q)$.

   \begin{figure}
   	\includegraphics[scale=1]{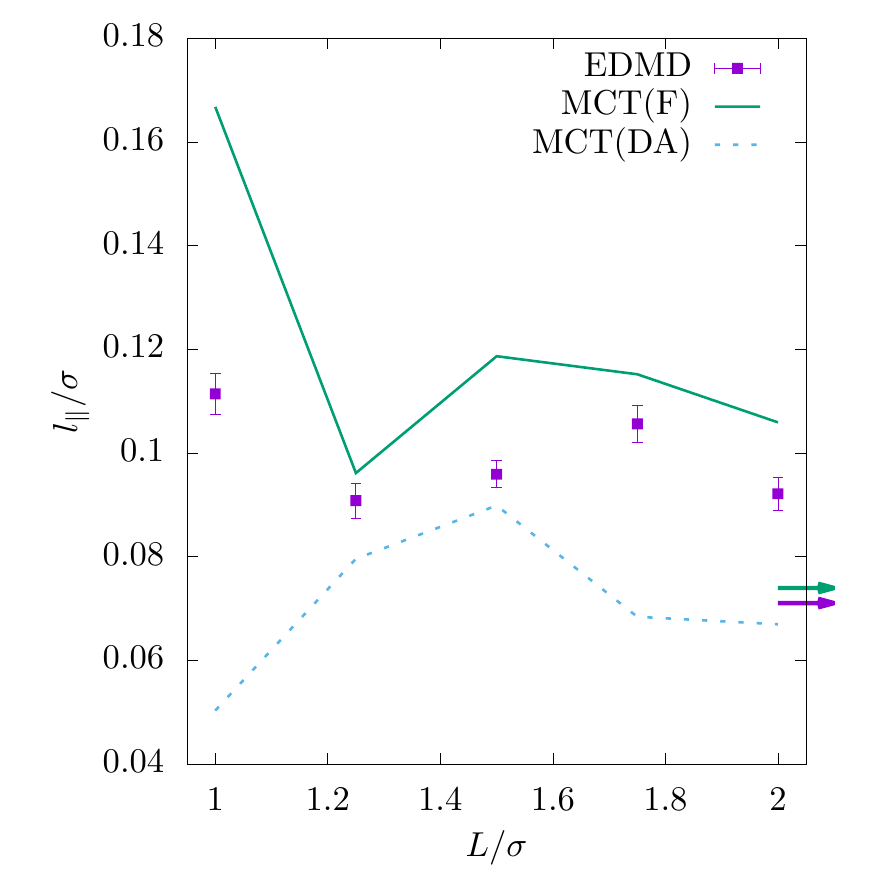}
	\caption{Localization length $ l_\parallel $ for different accessible wall separations $ L $ as determined from the von Schweidler fit Eq.~() of the mean{-}square displacement $ \delta r_\parallel^2{(t)} $ shown in Fig.~\ref{fig:msd}.  Results are shown for MCT using the full model (F) and the diagonal approximation (DA) and for even-driven molecular dynamics simulations (EDMD). Arrows indicate the bulk limit. }
	\label{fig:loc}
\end{figure}

Based on the in-plane {mean-square} displacement we have also extracted the localization length $l_\parallel$, which can be interpreted as the in-plane cage size. It shows an intricate non-monotonic behavior {highlighting} the changes in local packing {upon} changing the wall separation (see Fig.~\ref{fig:loc}). The most pronounced change is in fact the increase of $l_\parallel$ when reducing {$L\downarrow \sigma$}. This change is induced by the structural reformation from two layers with particle in-between  ($ L = 1.3\sigma$) to two separate layers {($ L \lesssim 1.2\sigma$)} \cite{Mandal2017a}. As discussed above, commensurate packing induces more {free} space for motion within the cages. This behavior is qualitatively predicted by the full solution of cMCT. The overestimation at $L=1.0\sigma$ is likely to be explained by polydispersity which reduces the effect of non-monotonic structural changes. It is noteworthy that in the case of the localization length $l_\parallel$ the diagonal approximation is not even qualitatively predicting the correct behavior. \gj{The DA also systematically underestimates the localization length which is directly related to the overestimation of the nonergodocity parameters observed in Figs.~\ref{fig:nonergo_coh} and \ref{fig:nonergo_incoh}. }
  
    \section{ Von Schweidler exponent}
  \label{sec:vonSchweidler}

     \begin{figure}
     	\includegraphics[scale=1]{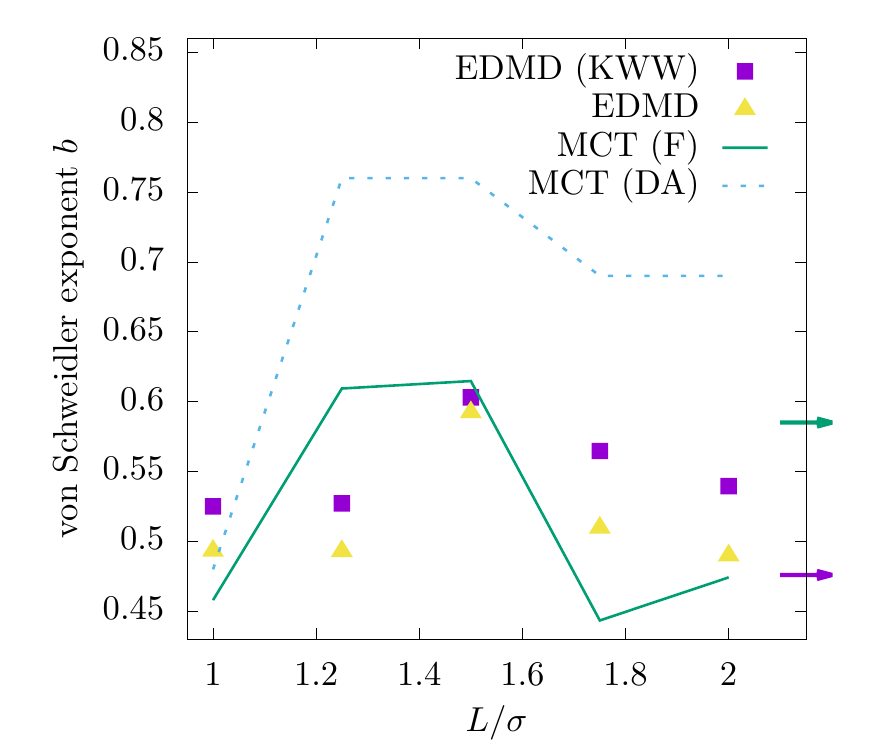}
  	\caption{Von Schweidler exponent $ b $ for various accessible slit widths $ L $. Results are shown for MCT using the full model (F) and the diagonal approximation (DA) and for even-driven molecular dynamics simulations (EDMD). The results are based on fitting the coherent ISF with Eq.~(\ref{eq:schweidler2}) with fixed nonergodicity parameters.  }
  	\label{fig:schweidler}
  \end{figure}

So far we have discussed one important feature of glassy dynamics: the formation of a plateau {indicating} structural arrest. However, also the decay from this plateau shows important properties specific for supercooled liquids. One of these features is the non-exponential decay. As has been discussed in Sec.~\ref{sec:struc_relax} and shown in Fig.~\ref{fig:susceptibility} the right flank of the low-frequency peak is described by the von Schweidler law, or late $\beta$ relaxation, with power exponent $b$ \cite{Gotze2009}. In fact, this power law exponent is related to the stretching exponent $\beta_\text{kww}$ in the KWW law (\ref{eq:KWW}) \cite{FUCHS1994241}. Both are nearly identical in our case and show the same non-monotonic behavior with the confinement length $L$ (see Fig.~\ref{fig:schweidler}). The simulations \gj{clearly indicate} that the stretching is strongest for commensurate packing. This behavior is quantitatively  predicted by the full solution of cMCT. 

\section{Conclusions}
\label{sec:conclusions}

While our previous theoretical results were based on an uncontrolled {technical} approximation we have introduced in this work a numerical procedure to solve the full matrix-valued system of integro-differential cMCT equations. We observe that the off-diagonal components, which introduce mode mixing, have an important impact on structural relaxation. Incorporating them into the numerical procedure is therefore crucial to obtain quantitative agreement with computer simulations. Nevertheless, the `diagonal approximation' is able to qualitatively predict most of the important features of confined dynamics. At least for the case of cMCT it can thus be concluded that the `diagonal approximation' is an appropriate numerical approximation to find stable solutions for MCT. This conclusion might be of relevance for other mode-coupling theories including active microrheology \cite{Gruber2019,PhysRevE.101.012612}, active particles \cite{liluashvili2017mode,reichert2021transport} or molecules \cite{schilling1997mode,PhysRevE.56.5659,PhysRevE.61.6934}. \gj{Conceptually, the same formalism could also be used to study confined hard discs as in Ref.~\cite{Yamchi2012_disks}, where we would expect to observe similar phenomenology.  However, the intrinsic mean-field nature of the mode-coupling approximation will become questionable in such strongly confined systems in low dimensions.}

Our work demonstrates that confinement {entails} non-trivially effects {for} the dynamical properties of glass-forming liquids. We have shown that mode-coupling theory in confinement can be used to make quantitative predictions for several important dynamical observables including the intermediate scattering function,  nonergodidcity parameters and von Schweidler exponents. The results can be rationalized using the violation of {translational} and rotational symmetries, leading to inhomogeneous and anisotropic dynamical response. We have also introduced the important concepts of commensurate and incommensurate packing which relate to non-monotonic structural reformation upon changing the confinement length $L$ and thus strongly impact the critical packing fractions and localization lengths. In future work it would be interesting to compare these detailed simulation and numerical results to the structural relaxation of colloidal liquids observed in confocal microscopy \cite{villada2022layering}.

 \section*{Acknowledgements} 
T.F. acknowledges support by the Austrian Science Fund (FWF): I 5257. The computational results presented have been achieved in part using the HPC infrastructure LEO of the University of Innsbruck.
 
 \FloatBarrier
 \appendix
 
 \section{Time-integration of the full MCT equations of motion}
 \label{app:integration}

 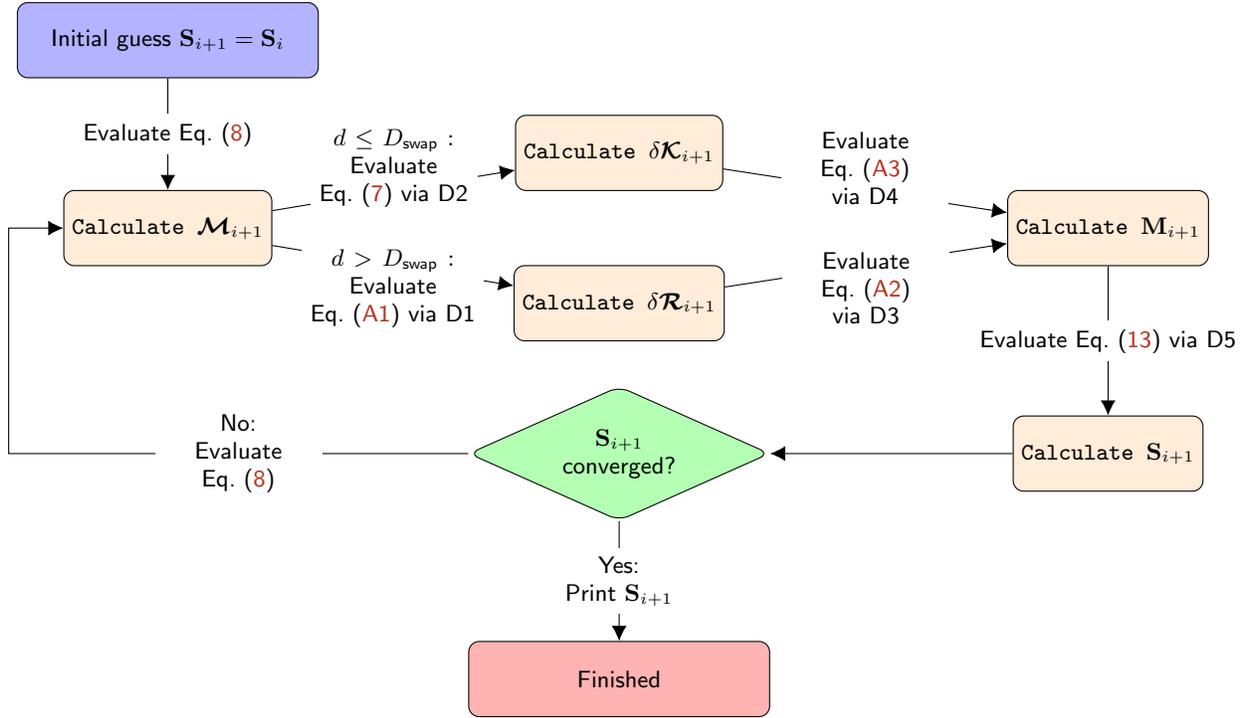
\begin{figure*}
 	\begin{tikzpicture}[node distance=1.5cm,
 	every node/.style={fill=white, font=\sffamily}, align=center]
 	\node (start)             [activityStarts]              {Initial guess $\mathbf{S}_{i+1}=\mathbf{S}_i$};
 	\node (M)     [process, below of=start, yshift=-1cm]          {Calculate $\bm{\mathcal{{M}}}_{i+1}$};
 	
 	\node (R)     [process, right of=M, yshift=-1cm, xshift=4.5cm]          {Calculate $\delta \bm{\mathcal{{R}}}_{i+1}$};
 	\node (K)     [process, right of=M, yshift=1cm, xshift=4.5cm]          {Calculate $\delta \bm{\mathcal{{K}}}_{i+1}$};
 	
 	\node (Meff)     [process, right of=M,  xshift=11.0cm]          {Calculate $\mathbf{M}_{i+1}$};
 	
 	\node (S)     [process, below of=Meff,  yshift=-1.5cm]          {Calculate $\mathbf{S}_{i+1}$};
 	
 	\node [test, left of=S, xshift=-5.0cm] (conv) {$\mathbf{S}_{i+1}$ converged?};
 	
 	\node [left of = conv, xshift=-6.5cm]  (c1)  {}; 
 	
 	\node (ActivityEnds)      [startstop, below of = conv, yshift = -1.5cm] {Finished};
 	
 	\draw[->]      (start) -- node[text width=4cm]
 	{Evaluate Eq.~(\ref{eq:MCT_functional})} (M);
 	
 	\draw[->]      (M) -- node[text width=2cm, yshift=0.3cm]
 	{ $d \leq D_\text{swap}:$  Evaluate Eq.~(\ref{eq:eom2}) via D2} (K);
 	
 	\draw[->]      (M) -- node[text width=2.2cm, yshift=-0.3cm]
 	{$d > D_\text{swap}:$ Evaluate Eq.~(\ref{M_to_Kr}) via D1} (R);
 	
 	 	\draw[->]      (K) -- node[text width=2cm, yshift=0.3cm]
 	{   Evaluate Eq.~(\ref{eq:K_to_Meff}) via D4} (Meff);
 	
 	\draw[->]      (R) -- node[text width=2cm, yshift=-0.3cm]
 	{ Evaluate Eq.~(\ref{eq:Kr_Meff}) via D3} (Meff);
 	
 	\draw[->]      (Meff) -- node[text width=4cm]
 	{Evaluate Eq.~(\ref{eq:eomS}) via D5} (S);
 	
 	\draw[->]      (S) -- (conv);
 	
 	\draw[->]      (conv)  -- node[text width=2cm]
 	{No:\\ Evaluate Eq.~(\ref{eq:MCT_functional})}  (c1.west)  |- (M);
 	
 	\draw[->]      (conv) -- node[text width=2cm]
 	{Yes: \\ Print $\mathbf{S}_{i+1}$} (ActivityEnds);
 	
 	 	\end{tikzpicture}
 	\caption{Flowchart for the time-integration of the MCT equations of motion. The flow chart visualizes Algorithm~\ref{alg:time_int}.}
 	\label{fig:flowchart}
 \end{figure*}

 \begin{figure}[t]
 	\begin{algorithm}[H]
 		\caption{Time-integration of the mode-coupling theory equations of motion, as sketched in Fig.~\ref{fig:flowchart}. The discretizations D1-D5 are described in the main text. }
 		\begin{algorithmic}[1]
 			\Input{ Time step size $ \Delta t $, decimation step $d$, tolerance $ tol = 10^{-6} $, discretized time dependence for the quantities, $ \bm{S}_i \leftarrow \bm{S}(i \Delta t) $,  $\bm{M}(i \Delta t) $,  $\bm{K}(i \Delta t)$,  $ \delta \bm{\mathcal{K}}(i \Delta t)$, $ \delta \bm{\mathcal{R}}(i \Delta t)$ and  $  \bm{\mathcal{M}}(i \Delta t)$.}
 			\Output { Updated time dependence $ \bm{S}_{i+1}$, $\bm{M}_{i+1}$,  $\bm{K}_{i+1}$,  $ \delta \bm{\mathcal{K}}_{i+1}$, $ \delta \bm{\mathcal{R}}_{i+1}$ and  $  \bm{\mathcal{M}}_{i+1}$ }
 			\State{Set $\bm{S}_{i+1} = \bm{S}_{i} $, $ \Delta = 1 $, $ \bm{S}^\text{old} = \bm{S}_{i+1} $ }
 			\While{$ \Delta > tol $}
 			
 			\State{ $\bm{\mathcal{M}}_{i+1}=\bm{\mathcal{F}}\left[\bm{S}_{i+1},\bm{S}_{i+1}\right]$ using Eq.~(\ref{eq:MCT_functional}).}
 			\If{$d \leq D_\text{swap}$}
 			\State{Calculate $\delta \bm{\mathcal{K}}_{i+1}$ using Eq.~(\ref{eq:eom2}) discretized with D3.}
 			\State{Calculate $ \mathbf{{K}}_{i+1}$ using the contraction Eq.~(\ref{eq:contraction}).}
 			\State{Calculate $ \bm{M}_{i+1}$ using Eq.~(\ref{eq:K_to_Meff}) with D4.}
 			\Else
 			\State{Calculate $\delta \bm{\mathcal{{R}}}_{i+1}$ using Eq.~(\ref{M_to_Kr}) with D1.}
 			\State{Calculate $\delta {\mathbf{{R}}}_{i+1}$ using the contraction Eq.~(\ref{eq:contraction}).}
 			\State{Calculate $ \bm{M}_{i+1}$ using Eq.~(\ref{eq:Kr_Meff}) with D2.}
 			\EndIf
 			\State{Calculate $\bm{S}_{i+1}$ using Eq.~(\ref{eq:eomS}) and $ \bm{M}_{i+1}$ discretized} \Statex{ \hspace*{0.415cm} with D5.}
 			\State{$\Delta = \max_{\mu, \nu} (|{S}_{\mu \nu, i+1} - {S}^\text{old}_{\mu \nu} |) $}
 			\State{$ \bm{S}^\text{old} = \bm{S}_{i+1} $}
 			
 			\EndWhile
 		\end{algorithmic}
 		\label{alg:time_int}
 	\end{algorithm}
 \end{figure}

 Our algorithm, which is visualized in the flow chart in Fig.~\ref{fig:flowchart} and described in more detail in Algorithm \ref{alg:time_int}, is a combination of a short-time integration via route A, followed by a long-time integration via route B. The algorithm features five different discretizations of differential equations, D1-D5, which will be presented in the following.
 \begin{enumerate}[label=D\arabic*]
 	\item describes the discretization in time of the equation of motion for the generalized mean-square displacement Eq.~(\ref{eq:msd_general}). To integrate it we take the first time-derivative,
 	\begin{equation}\label{M_to_Kr}
 	 \delta \bm{\mathcal{\dot{R}}}(t) + \bm{\mathcal{D}} \frac{\text{d}}{\text{d} t} \int_0^t  \bm{\mathcal{{M}}}(t-t') \delta \bm{\mathcal{{R}}}(t') \text{d}t' = \bm{\mathcal{D}} .
 	\end{equation}
 	 This equation is then discretized in the same way as the mean-square displacement described in Ref.~\cite{Jung2020_self} Eq.~(A4) and Eq.~(A2) using the algorithm 'Integro-differential method with moments' presented in Chapter 3.5.1. of Ref.~\cite{Gruber2019}.
 	  	\item denotes the direct calculation of the force kernel $\delta \bm{\mathcal{K}}_{i+1}$ via Eq.~(\ref{eq:eom2}) for route A. This equation has the same form as Eq.~(\ref{eq:msd_general}) and thus the algorithm described in D1 can be applied. (Including the discretization of the additional external driving term $\bm{\mathcal{D}} \bm{\mathcal{M}} (t) \bm{\mathcal{D}}$ as described in Chapter 3.5.1. of Ref.~\cite{Gruber2019}.)
 	\item describes the calculation of the effective memory kernel via ``route B'', using the contracted generalized mean-square displacement $ \delta {\ddot{R}}_{\mu \nu}(t) = {\delta {{K}_{\mu \nu}}}(t) $, which in time domain can be derived from Eq.~(\ref{eq:def_meff}). We find,
 	\begin{equation}\label{eq:Kr_Meff}
 	 \delta \mathbf{\dot{R}}(t) + \mathbf{D} \frac{\text{d}}{\text{d} t} \int_{0}^{t} \text{d}s \,\mathbf{M}(s)\delta \mathbf{{R}}(t-s) = \mathbf{{D}},
 	\end{equation}
 	using the relations, $\delta \mathbf{\dot{{{R}}}}(0) = \mathcal{C} \{ \bm{\dot{\mathcal{{R}}}}(0) \} = \mathbf{{D}} $ and $\delta \mathbf{{{{R}}}}(0) = \mathcal{C} \{ \delta \bm{{\mathcal{{R}}}}(0) \} = 0 $.
 	
 	The equation (\ref{eq:Kr_Meff}) can be discretized in the same way as described in D1 above. In this special case, the discretized equation of motion is then solved for the effective memory kernel $ \mathbf{M}_{i+1} = \mathbf{M}\big((t+1)\Delta t\big) $, where $ \Delta t $ describes the time step size. 
 	
 	\item describes the evaluation of the effective memory kernel in route A and is given in time domain by the equation,
 	\begin{equation}\label{eq:K_to_Meff}
 	\mathbf{M}(t) + \mathbf{D}^{-1} \delta \mathbf{K}(t)\mathbf{D}^{-1}    + \int_{0}^{t} \text{d}s\, \mathbf{M}(s)\delta \mathbf{K}(t-s)\mathbf{D}^{-1} = 0.
 	\end{equation}
 	This equation also has the same form as Eq.~(\ref{eq:msd_general}) and thus the algorithm described in D1 can be applied. Also in this case the final discritized equation is solved for $ \mathbf{M}_{i+1}. $
 	\item describes the final integration to obtain $ \mathbf{S}_{i+1} $ in Eq.~(\ref{eq:eomS}). This equation is equivalent to Eq.~(42) in Ref.~\cite{Jung2020_self} and its discretization is given in Eq.~(A1) of Ref.~\cite{Jung2020_self}. 
 \end{enumerate}
 
 To enable the numerical integration of these equations over several orders of magnitude in time, we employ the decimation procedure discussed extensively in the literature \cite{Sperl2000,Gruber2019}. The idea is to use an adaptive time step $ \Delta t $ which is doubled every $ N_t/2 $ discretization steps. In our specific implementation we start with $ N_t = 1024 $, but skip a decimation at $ d=  22,24,31$ and $36$, meaning that the number of discretization steps $ N_t/2 $ per decimation is doubled and peaks at $N_t=16385$ after 36 decimations. This enables a stable integration up to times {$ t \gtrsim 10^5 \tau$}, starting from {$ \Delta t = 10^{-9}\tau. $} The decimation step at which ``route B'' replaces ``route A'' is set to $ D_\text{swap} = 18$. 
 
 \section{Finite dimensionality and numerical evaluation}
 \label{sec:intro_numerics}
 
 The evaluation of the mode-coupling functional requires a discretization into a finite set of wavenumbers {$q= q_0,\ldots, q_{N-1}$} and the mode indices need to be restricted $ |\mu| \leq M. $ Similar to Refs.~\cite{Jung:2020,Jung2020_self} we found empirically $ N=30 $ and $ M=5 $ for converging solutions. This discretization, however, leads to an inconsistency in the model. Generally the direct correlation function and the structure factor are connected via the Ornstein-Zernike (OZ) equation,
 \begin{equation}\label{key}
 \bm{S}^{-1}(q) = \frac{n_0}{L^2} \left[ \mathbf{v} - \mathbf{c}(q)  \right],
 \end{equation} 
 with $ [\mathbf{v}]_{\mu \nu} = v_{\mu - \nu} $. {Upon} introducing the cutoff $ |\mu| \leq M $ one has to decide whether the quantity $ \bm{S}^{-1}(q) $ should be calculated using the OZ equation with the finite dimensional $ \mathbf{c}(q) $ or by inverting the finite dimensional $ \bm{S}(q). $ In general both choices will yield different results. Here, we have chosen the second variant, since otherwise the solution from the iteration for the nonergodicity parameter and the full time integration might yield slightly different values. We have indeed checked the consistency and validated that our choice indeed implies that both solutions are consistent. It should be emphasized that these discretization errors in the full model can be systematically reduced by increasing $ M $ which stands in stark contrast to the above introduced diagonal approximation which is uncontrolled.

 The algorithm to calculate the nonergodicity parameter described in Section~\ref{sec:intro_nonergo} allows {locating} the critical packing fraction $ \varphi_c $ in a numerical procedure with an accuracy of roughly $\epsilon = (\varphi - \varphi_c)/\varphi_c \approx \pm 10^{-5}  $. When attempting to approach the critical point even closer we observe the emergence of non-monotonicities and instabilities in the iteration algorithm which have to be caused by the cutoff $ M $ for the mode indices, because they violate the theorems derived in Ref.~\cite{Lang2012}. Despite these small instabilities induced by the finite-dimensional matrices and their inversion the accuracy is sufficient to determine reliable values for the critical packing fraction $ \varphi_c $ and the nonergodicity parameter $ \bm{F}_{\mu \nu} $. 
 
 To achieve stable solutions for the full time\delete{-}dependence of the incoherent and coherent intermediate scattering function as introduced in Section~\ref{sec:intro_time} up to times $ t \approx 10^5 \tau $ a fine discretization is required. Therefore, a full run consists of over $10^5$ individual time steps, each requiring several times the evaluation of the mode-coupling functional defined in Eqs.~(\ref{eq:MCT_functional}) and (\ref{eq:vertices}), which is clearly the bottleneck of the integration algorithm. 

The time integration described above scales as $ \mathcal{O}(N_t N M^3) $, where {$N_t$ denotes the number of time steps,}  $ N $ defines the total amount of wavenumbers considered (``$ q $-discretization'')\add{,} and $ M $ the highest mode, $ |\mu|\leq M. $ The scaling $  M^3 $ originates from the matrix-matrix multiplications which have to be evaluated to solve these equations.

The evaluation of the mode-coupling functional from the scattering functions as defined in Eqs.~(\ref{eq:MCT_functional}) and (\ref{eq:vertices}) scales as $ \mathcal{O}(N^3 M^7) $ and is thus clearly the bottleneck of the evaluation. In the following we show how to precalculate the inner sums to reduce the scaling to $ \mathcal{O}(N^3 M^4) $, which effectively speeds up the numerics by a factor of 1000 for $ M=5 $, which is the value used in this work.

First, the sum over $ \kappa $ in the vertices in Eq.~(\ref{eq:vertices}) can be separated from the rest of the calculation since they are independent of $ q $. When the final kernel is evaluated, the inclusion of these sums requires just a matrix-matrix multiplication scaling as $ \mathcal{O}(N M^3) $. Second,
when inserting the vertices {in} Eq.~(\ref{eq:vertices}) into the functional, the inner sum in the vertices can be expanded. The most complex term that {arises} in this decomposition is given by
\begin{align}\label{key}
\prescript{1}{}T_{\mu \nu}^{\perp \perp}(q) &= \frac{n_0^2}{2N L^4} \sum_{\substack{\bm{q}_1,\\\bm{q}_2=\bm{q}-\bm{q}_1}}  \sum_{\mu_1,\nu_1} \\& \times \left(\sum_{\mu_1} Q_{\mu-\mu_1} c_{\mu_2,\mu-\mu_1}(q_2) S_{\mu_1\nu_1}(q_1,t) \right)  \\ & \times \left(  \sum_{\nu_2} Q_{\nu-\nu_2} c_{\nu_1,\nu-\nu_2}(q_1) S_{\mu_2\nu_2}(q_2,t) \right) \add{.}
\end{align}
The inner sums can be calculated with a time complexity of $ \mathcal{O}(N^2M^5) $. The above restructuring of the total sum is possible for all different terms of the decomposition and all combinations of $ \alpha=(\parallel,\perp) $ and $ \beta=(\parallel,\perp) $. When calculating the MCT functional for the incoherent scattering function the same decomposition applies, only that some of the sums have to be specifically evaluated including the incoherent intermediate scattering function. In conclusion, precalculation of these inner sums allows {evaluating} the mode-coupling functional in $ \mathcal{O}(N^3 M^4) $. The final algorithm was additionally parallelized using OpenMP, enabling the full integration of the equations of motion up to times $ t \approx 10^5 \tau $ on 8 cores on a standard CPU in roughly 1-2 weeks.

 \section{Von\add{-}Schweidler exponent and asymptotic expansion}

In this manuscript we have calculated the von\add{-}Schweidler exponent $ b $ by fitting the $ \beta $-scaling equation defined in Eq.~(\ref{eq:schweidler}) for every value of the wavenumber $ q $ and mode indices $ \mu,\nu $ to the coherent scattering function $ S_{\mu \nu}(q,t). $ The values for $ F_{\mu \nu}(q) $ were fixed using the solution from the iteration described in Section~\ref{sec:intro_nonergo}.

The von Schweidler exponent could, in principle, also be directly determined from the nonergodicity parameters using the asymptotic expansion derived in Ref.~\cite{Jung:2020_scalingtheory}. The asymptotic expansion is based on the assumption that the glass transition in mode-coupling theory arises as a bifurcation transition. This allows {determining} numerically a characteristic eigenvector from an eigenvalue equation which can be evaluated to finally calculate the von\add{-}Schweidler exponent $ b $ \cite{Franosch1997,Jung:2020_scalingtheory}. The validity of these asymptotic formulas has been confirmed in Ref.~\cite{Jung:2020} for the diagonal approximation. 

When applying this expansion to the results obtained in the present manuscript for the full model we observe that the predicted exponents from the asymptotic theory deviate by roughly 5-10\% from the ones determined using the fitting procedure described above. We believe that this deviation {arises} from the numerical instabilities in the iteration procedure when approaching the critical packing fraction $ \varphi_c $ as described in Section~\ref{sec:intro_numerics}. Our hypothesis is based on the observation that the eigenvalue {corresponding} to the characteristic eigenvector is not precisely 1 as would be assumed directly at the bifurcation transition. We attempted to systematically correct the nonergodicity parameters and observed that marginal changes, corresponding to less than 0.1\% of the absolute values, induced significant changes in the von Schweidler exponent $ b $ although the eigenvalue only barely changed. This shows that the expansion critically depends on very precise values for the nonergodicity parameter. Based on this analysis we decided to report only the exponents calculated by the fitting procedure of the full time dependence of the coherent scattering function. We have therefore decided to not report results of the asymptotic expansion.

\bibliography{library_local.bib}

\begin{thebibliography}{87}%
\makeatletter
\providecommand \@ifxundefined [1]{%
 \@ifx{#1\undefined}
}%
\providecommand \@ifnum [1]{%
 \ifnum #1\expandafter \@firstoftwo
 \else \expandafter \@secondoftwo
 \fi
}%
\providecommand \@ifx [1]{%
 \ifx #1\expandafter \@firstoftwo
 \else \expandafter \@secondoftwo
 \fi
}%
\providecommand \natexlab [1]{#1}%
\providecommand \enquote  [1]{``#1''}%
\providecommand \bibnamefont  [1]{#1}%
\providecommand \bibfnamefont [1]{#1}%
\providecommand \citenamefont [1]{#1}%
\providecommand \href@noop [0]{\@secondoftwo}%
\providecommand \href [0]{\begingroup \@sanitize@url \@href}%
\providecommand \@href[1]{\@@startlink{#1}\@@href}%
\providecommand \@@href[1]{\endgroup#1\@@endlink}%
\providecommand \@sanitize@url [0]{\catcode `\\12\catcode `\$12\catcode
  `\&12\catcode `\#12\catcode `\^12\catcode `\_12\catcode `\%12\relax}%
\providecommand \@@startlink[1]{}%
\providecommand \@@endlink[0]{}%
\providecommand \url  [0]{\begingroup\@sanitize@url \@url }%
\providecommand \@url [1]{\endgroup\@href {#1}{\urlprefix }}%
\providecommand \urlprefix  [0]{URL }%
\providecommand \Eprint [0]{\href }%
\providecommand \doibase [0]{https://doi.org/}%
\providecommand \selectlanguage [0]{\@gobble}%
\providecommand \bibinfo  [0]{\@secondoftwo}%
\providecommand \bibfield  [0]{\@secondoftwo}%
\providecommand \translation [1]{[#1]}%
\providecommand \BibitemOpen [0]{}%
\providecommand \bibitemStop [0]{}%
\providecommand \bibitemNoStop [0]{.\EOS\space}%
\providecommand \EOS [0]{\spacefactor3000\relax}%
\providecommand \BibitemShut  [1]{\csname bibitem#1\endcsname}%
\let\auto@bib@innerbib\@empty
\bibitem [{\citenamefont {González‐Mozuelos}\ \emph
  {et~al.}(1991)\citenamefont {González‐Mozuelos}, \citenamefont
  {Alejandre},\ and\ \citenamefont {Medina‐Noyola}}]{Noyola1991}%
  \BibitemOpen
  \bibfield  {author} {\bibinfo {author} {\bibfnamefont {P.}~\bibnamefont
  {González‐Mozuelos}}, \bibinfo {author} {\bibfnamefont {J.}~\bibnamefont
  {Alejandre}},\ and\ \bibinfo {author} {\bibfnamefont {M.}~\bibnamefont
  {Medina‐Noyola}},\ }\bibfield  {title} {\bibinfo {title} {Structure of a
  colloidal suspension confined in a planar slit},\ }\href
  {https://doi.org/10.1063/1.461260} {\bibfield  {journal} {\bibinfo  {journal}
  {The Journal of Chemical Physics}\ }\textbf {\bibinfo {volume} {95}},\
  \bibinfo {pages} {8337} (\bibinfo {year} {1991})}\BibitemShut {NoStop}%
\bibitem [{\citenamefont {N\'emeth}\ and\ \citenamefont
  {L\"owen}(1999)}]{Nemeth:PRE59:1999}%
  \BibitemOpen
  \bibfield  {author} {\bibinfo {author} {\bibfnamefont {Z.~T.}\ \bibnamefont
  {N\'emeth}}\ and\ \bibinfo {author} {\bibfnamefont {H.}~\bibnamefont
  {L\"owen}},\ }\bibfield  {title} {\bibinfo {title} {Freezing and glass
  transition of hard spheres in cavities},\ }\href
  {https://doi.org/10.1103/PhysRevE.59.6824} {\bibfield  {journal} {\bibinfo
  {journal} {Phys. Rev. E}\ }\textbf {\bibinfo {volume} {59}},\ \bibinfo
  {pages} {6824} (\bibinfo {year} {1999})}\BibitemShut {NoStop}%
\bibitem [{\citenamefont {Nygård}(2017)}]{Nygard2017}%
  \BibitemOpen
  \bibfield  {author} {\bibinfo {author} {\bibfnamefont {K.}~\bibnamefont
  {Nygård}},\ }\bibfield  {title} {\bibinfo {title} {Colloidal diffusion in
  confined geometries},\ }\href {https://doi.org/10.1039/C7CP02497E} {\bibfield
   {journal} {\bibinfo  {journal} {Phys. Chem. Chem. Phys.}\ }\textbf {\bibinfo
  {volume} {19}},\ \bibinfo {pages} {23632} (\bibinfo {year}
  {2017})}\BibitemShut {NoStop}%
\bibitem [{\citenamefont {Granick}(1991)}]{Granick1991MotionsAR}%
  \BibitemOpen
  \bibfield  {author} {\bibinfo {author} {\bibfnamefont {S.}~\bibnamefont
  {Granick}},\ }\bibfield  {title} {\bibinfo {title} {Motions and relaxations
  of confined liquids.},\ }\href@noop {} {\bibfield  {journal} {\bibinfo
  {journal} {Science}\ }\textbf {\bibinfo {volume} {253 5026}},\ \bibinfo
  {pages} {1374} (\bibinfo {year} {1991})}\BibitemShut {NoStop}%
\bibitem [{\citenamefont {Mittal}\ \emph {et~al.}(2008)\citenamefont {Mittal},
  \citenamefont {Truskett}, \citenamefont {Errington},\ and\ \citenamefont
  {Hummer}}]{Mittal:PRL100:2008}%
  \BibitemOpen
  \bibfield  {author} {\bibinfo {author} {\bibfnamefont {J.}~\bibnamefont
  {Mittal}}, \bibinfo {author} {\bibfnamefont {T.~M.}\ \bibnamefont
  {Truskett}}, \bibinfo {author} {\bibfnamefont {J.~R.}\ \bibnamefont
  {Errington}},\ and\ \bibinfo {author} {\bibfnamefont {G.}~\bibnamefont
  {Hummer}},\ }\bibfield  {title} {\bibinfo {title} {Layering and
  {position}-{dependent} {diffusive} {dynamics} of {confined} {fluids}},\
  }\href {https://doi.org/10.1103/PhysRevLett.100.145901} {\bibfield  {journal}
  {\bibinfo  {journal} {Phys. Rev. Lett.}\ }\textbf {\bibinfo {volume} {100}},\
  \bibinfo {pages} {145901} (\bibinfo {year} {2008})}\BibitemShut {NoStop}%
\bibitem [{\citenamefont {Pieranski}\ \emph {et~al.}(1983)\citenamefont
  {Pieranski}, \citenamefont {Strzelecki},\ and\ \citenamefont
  {Pansu}}]{exp:Pieranski1983}%
  \BibitemOpen
  \bibfield  {author} {\bibinfo {author} {\bibfnamefont {P.}~\bibnamefont
  {Pieranski}}, \bibinfo {author} {\bibfnamefont {L.}~\bibnamefont
  {Strzelecki}},\ and\ \bibinfo {author} {\bibfnamefont {B.}~\bibnamefont
  {Pansu}},\ }\bibfield  {title} {\bibinfo {title} {Thin colloidal crystals},\
  }\href {https://doi.org/10.1103/PhysRevLett.50.900} {\bibfield  {journal}
  {\bibinfo  {journal} {Phys. Rev. Lett.}\ }\textbf {\bibinfo {volume} {50}},\
  \bibinfo {pages} {900} (\bibinfo {year} {1983})}\BibitemShut {NoStop}%
\bibitem [{\citenamefont {Schmidt}\ and\ \citenamefont
  {L\"owen}(1996)}]{theo:Schmidt1996}%
  \BibitemOpen
  \bibfield  {author} {\bibinfo {author} {\bibfnamefont {M.}~\bibnamefont
  {Schmidt}}\ and\ \bibinfo {author} {\bibfnamefont {H.}~\bibnamefont
  {L\"owen}},\ }\bibfield  {title} {\bibinfo {title} {Freezing between two and
  three dimensions},\ }\href {https://doi.org/10.1103/PhysRevLett.76.4552}
  {\bibfield  {journal} {\bibinfo  {journal} {Phys. Rev. Lett.}\ }\textbf
  {\bibinfo {volume} {76}},\ \bibinfo {pages} {4552} (\bibinfo {year}
  {1996})}\BibitemShut {NoStop}%
\bibitem [{\citenamefont {Schmidt}\ and\ \citenamefont
  {L\"owen}(1997)}]{theo:Schmidt1997}%
  \BibitemOpen
  \bibfield  {author} {\bibinfo {author} {\bibfnamefont {M.}~\bibnamefont
  {Schmidt}}\ and\ \bibinfo {author} {\bibfnamefont {H.}~\bibnamefont
  {L\"owen}},\ }\bibfield  {title} {\bibinfo {title} {Phase diagram of hard
  spheres confined between two parallel plates},\ }\href
  {https://doi.org/10.1103/PhysRevE.55.7228} {\bibfield  {journal} {\bibinfo
  {journal} {Phys. Rev. E}\ }\textbf {\bibinfo {volume} {55}},\ \bibinfo
  {pages} {7228} (\bibinfo {year} {1997})}\BibitemShut {NoStop}%
\bibitem [{\citenamefont {Alba-Simionesco}\ \emph {et~al.}(2006)\citenamefont
  {Alba-Simionesco}, \citenamefont {Coasne}, \citenamefont {Dosseh},
  \citenamefont {Dudziak}, \citenamefont {Gubbins}, \citenamefont
  {Radhakrishnan},\ and\ \citenamefont
  {Sliwinska-Bartkowiak}}]{Alba-Simionesco_2006}%
  \BibitemOpen
  \bibfield  {author} {\bibinfo {author} {\bibfnamefont {C.}~\bibnamefont
  {Alba-Simionesco}}, \bibinfo {author} {\bibfnamefont {B.}~\bibnamefont
  {Coasne}}, \bibinfo {author} {\bibfnamefont {G.}~\bibnamefont {Dosseh}},
  \bibinfo {author} {\bibfnamefont {G.}~\bibnamefont {Dudziak}}, \bibinfo
  {author} {\bibfnamefont {K.~E.}\ \bibnamefont {Gubbins}}, \bibinfo {author}
  {\bibfnamefont {R.}~\bibnamefont {Radhakrishnan}},\ and\ \bibinfo {author}
  {\bibfnamefont {M.}~\bibnamefont {Sliwinska-Bartkowiak}},\ }\bibfield
  {title} {\bibinfo {title} {Effects of confinement on freezing and melting},\
  }\href {https://doi.org/10.1088/0953-8984/18/6/R01} {\bibfield  {journal}
  {\bibinfo  {journal} {Journal of Physics: Condensed Matter}\ }\textbf
  {\bibinfo {volume} {18}},\ \bibinfo {pages} {R15} (\bibinfo {year}
  {2006})}\BibitemShut {NoStop}%
\bibitem [{\citenamefont {Mandal}\ \emph {et~al.}(2014)\citenamefont {Mandal},
  \citenamefont {Lang}, \citenamefont {Gross}, \citenamefont {Oettel},
  \citenamefont {Raabe}, \citenamefont {Franosch},\ and\ \citenamefont
  {Varnik}}]{Mandal2014}%
  \BibitemOpen
  \bibfield  {author} {\bibinfo {author} {\bibfnamefont {S.}~\bibnamefont
  {Mandal}}, \bibinfo {author} {\bibfnamefont {S.}~\bibnamefont {Lang}},
  \bibinfo {author} {\bibfnamefont {M.}~\bibnamefont {Gross}}, \bibinfo
  {author} {\bibfnamefont {M.}~\bibnamefont {Oettel}}, \bibinfo {author}
  {\bibfnamefont {D.}~\bibnamefont {Raabe}}, \bibinfo {author} {\bibfnamefont
  {T.}~\bibnamefont {Franosch}},\ and\ \bibinfo {author} {\bibfnamefont
  {F.}~\bibnamefont {Varnik}},\ }\bibfield  {title} {\bibinfo {title} {Multiple
  reentrant glass transitions in confined hard-sphere glasses},\ }\href
  {https://doi.org/10.1038/ncomms5435} {\bibfield  {journal} {\bibinfo
  {journal} {Nat. Commun.}\ }\textbf {\bibinfo {volume} {5}},\ \bibinfo {pages}
  {4435} (\bibinfo {year} {2014})}\BibitemShut {NoStop}%
\bibitem [{\citenamefont {Sarangapani}\ \emph {et~al.}(2012)\citenamefont
  {Sarangapani}, \citenamefont {Schofield},\ and\ \citenamefont
  {Zhu}}]{C1SM06502E}%
  \BibitemOpen
  \bibfield  {author} {\bibinfo {author} {\bibfnamefont {P.~S.}\ \bibnamefont
  {Sarangapani}}, \bibinfo {author} {\bibfnamefont {A.~B.}\ \bibnamefont
  {Schofield}},\ and\ \bibinfo {author} {\bibfnamefont {Y.}~\bibnamefont
  {Zhu}},\ }\bibfield  {title} {\bibinfo {title} {Relationship between
  cooperative motion and the confinement length scale in confined colloidal
  liquids},\ }\href {https://doi.org/10.1039/C1SM06502E} {\bibfield  {journal}
  {\bibinfo  {journal} {Soft Matter}\ }\textbf {\bibinfo {volume} {8}},\
  \bibinfo {pages} {814} (\bibinfo {year} {2012})}\BibitemShut {NoStop}%
\bibitem [{\citenamefont {Sarangapani}\ \emph {et~al.}(2011)\citenamefont
  {Sarangapani}, \citenamefont {Schofield},\ and\ \citenamefont
  {Zhu}}]{PhysRevE.83.030502}%
  \BibitemOpen
  \bibfield  {author} {\bibinfo {author} {\bibfnamefont {P.~S.}\ \bibnamefont
  {Sarangapani}}, \bibinfo {author} {\bibfnamefont {A.~B.}\ \bibnamefont
  {Schofield}},\ and\ \bibinfo {author} {\bibfnamefont {Y.}~\bibnamefont
  {Zhu}},\ }\bibfield  {title} {\bibinfo {title} {Direct experimental evidence
  of growing dynamic length scales in confined colloidal liquids},\ }\href
  {https://doi.org/10.1103/PhysRevE.83.030502} {\bibfield  {journal} {\bibinfo
  {journal} {Phys. Rev. E}\ }\textbf {\bibinfo {volume} {83}},\ \bibinfo
  {pages} {030502(R)} (\bibinfo {year} {2011})}\BibitemShut {NoStop}%
\bibitem [{\citenamefont {Nugent}\ \emph {et~al.}(2007)\citenamefont {Nugent},
  \citenamefont {Edmond}, \citenamefont {Patel},\ and\ \citenamefont
  {Weeks}}]{PhysRevLett.99.025702}%
  \BibitemOpen
  \bibfield  {author} {\bibinfo {author} {\bibfnamefont {C.~R.}\ \bibnamefont
  {Nugent}}, \bibinfo {author} {\bibfnamefont {K.~V.}\ \bibnamefont {Edmond}},
  \bibinfo {author} {\bibfnamefont {H.~N.}\ \bibnamefont {Patel}},\ and\
  \bibinfo {author} {\bibfnamefont {E.~R.}\ \bibnamefont {Weeks}},\ }\bibfield
  {title} {\bibinfo {title} {Colloidal {glass} {transition} {observed} in
  {confinement}},\ }\href {https://doi.org/10.1103/PhysRevLett.99.025702}
  {\bibfield  {journal} {\bibinfo  {journal} {Phys. Rev. Lett.}\ }\textbf
  {\bibinfo {volume} {99}},\ \bibinfo {pages} {025702} (\bibinfo {year}
  {2007})}\BibitemShut {NoStop}%
\bibitem [{\citenamefont {Hunter}\ \emph {et~al.}(2014)\citenamefont {Hunter},
  \citenamefont {Edmond},\ and\ \citenamefont
  {Weeks}}]{PhysRevLett.112.218302}%
  \BibitemOpen
  \bibfield  {author} {\bibinfo {author} {\bibfnamefont {G.~L.}\ \bibnamefont
  {Hunter}}, \bibinfo {author} {\bibfnamefont {K.~V.}\ \bibnamefont {Edmond}},\
  and\ \bibinfo {author} {\bibfnamefont {E.~R.}\ \bibnamefont {Weeks}},\
  }\bibfield  {title} {\bibinfo {title} {Boundary {mobility} {controls}
  {glassiness} in {confined} {colloidal} {liquids}},\ }\href
  {https://doi.org/10.1103/PhysRevLett.112.218302} {\bibfield  {journal}
  {\bibinfo  {journal} {Phys. Rev. Lett.}\ }\textbf {\bibinfo {volume} {112}},\
  \bibinfo {pages} {218302} (\bibinfo {year} {2014})}\BibitemShut {NoStop}%
\bibitem [{\citenamefont {Williams}\ \emph {et~al.}(2015)\citenamefont
  {Williams}, \citenamefont {O\ifmmode~\breve{g}\else \u{g}\fi{}uz},
  \citenamefont {Bartlett}, \citenamefont {L\"owen},\ and\ \citenamefont
  {Patrick~Royall}}]{doi:10.1063/1.4905472}%
  \BibitemOpen
  \bibfield  {author} {\bibinfo {author} {\bibfnamefont {I.}~\bibnamefont
  {Williams}}, \bibinfo {author} {\bibfnamefont {E.~C.}\ \bibnamefont
  {O\ifmmode~\breve{g}\else \u{g}\fi{}uz}}, \bibinfo {author} {\bibfnamefont
  {P.}~\bibnamefont {Bartlett}}, \bibinfo {author} {\bibfnamefont
  {H.}~\bibnamefont {L\"owen}},\ and\ \bibinfo {author} {\bibfnamefont
  {C.}~\bibnamefont {Patrick~Royall}},\ }\bibfield  {title} {\bibinfo {title}
  {Flexible confinement leads to multiple relaxation regimes in glassy
  colloidal liquids},\ }\href {https://doi.org/10.1063/1.4905472} {\bibfield
  {journal} {\bibinfo  {journal} {J. Chem. Phys.}\ }\textbf {\bibinfo {volume}
  {142}},\ \bibinfo {pages} {024505} (\bibinfo {year} {2015})}\BibitemShut
  {NoStop}%
\bibitem [{\citenamefont {Nyg\aa{}rd}\ \emph {et~al.}(2012)\citenamefont
  {Nyg\aa{}rd}, \citenamefont {Kjellander}, \citenamefont {Sarman},
  \citenamefont {Chodankar}, \citenamefont {Perret}, \citenamefont
  {Buitenhuis},\ and\ \citenamefont {van~der Veen}}]{PhysRevLett.108.037802}%
  \BibitemOpen
  \bibfield  {author} {\bibinfo {author} {\bibfnamefont {K.}~\bibnamefont
  {Nyg\aa{}rd}}, \bibinfo {author} {\bibfnamefont {R.}~\bibnamefont
  {Kjellander}}, \bibinfo {author} {\bibfnamefont {S.}~\bibnamefont {Sarman}},
  \bibinfo {author} {\bibfnamefont {S.}~\bibnamefont {Chodankar}}, \bibinfo
  {author} {\bibfnamefont {E.}~\bibnamefont {Perret}}, \bibinfo {author}
  {\bibfnamefont {J.}~\bibnamefont {Buitenhuis}},\ and\ \bibinfo {author}
  {\bibfnamefont {J.~F.}\ \bibnamefont {van~der Veen}},\ }\bibfield  {title}
  {\bibinfo {title} {Anisotropic {pair} {correlations} and {structure}
  {factors} of {confined} {hard}-{sphere} {fluids}: An {experimental} and
  {theoretical} {study}},\ }\href
  {https://doi.org/10.1103/PhysRevLett.108.037802} {\bibfield  {journal}
  {\bibinfo  {journal} {Phys. Rev. Lett.}\ }\textbf {\bibinfo {volume} {108}},\
  \bibinfo {pages} {037802} (\bibinfo {year} {2012})}\BibitemShut {NoStop}%
\bibitem [{\citenamefont {Nyg\aa{}rd}\ \emph {et~al.}(2013)\citenamefont
  {Nyg\aa{}rd}, \citenamefont {Sarman},\ and\ \citenamefont
  {Kjellander}}]{doi:10.1063/1.4825176}%
  \BibitemOpen
  \bibfield  {author} {\bibinfo {author} {\bibfnamefont {K.}~\bibnamefont
  {Nyg\aa{}rd}}, \bibinfo {author} {\bibfnamefont {S.}~\bibnamefont {Sarman}},\
  and\ \bibinfo {author} {\bibfnamefont {R.}~\bibnamefont {Kjellander}},\
  }\bibfield  {title} {\bibinfo {title} {Local order variations in confined
  hard-sphere fluids},\ }\href {https://doi.org/10.1063/1.4825176} {\bibfield
  {journal} {\bibinfo  {journal} {J. Chem. Phys.}\ }\textbf {\bibinfo {volume}
  {139}},\ \bibinfo {pages} {164701} (\bibinfo {year} {2013})}\BibitemShut
  {NoStop}%
\bibitem [{\citenamefont {Nyg\aa{}rd}\ \emph
  {et~al.}(2016{\natexlab{a}})\citenamefont {Nyg\aa{}rd}, \citenamefont
  {Buitenhuis}, \citenamefont {Kagias}, \citenamefont {Jefimovs}, \citenamefont
  {Zontone},\ and\ \citenamefont {Chushkin}}]{PhysRevLett.116.167801}%
  \BibitemOpen
  \bibfield  {author} {\bibinfo {author} {\bibfnamefont {K.}~\bibnamefont
  {Nyg\aa{}rd}}, \bibinfo {author} {\bibfnamefont {J.}~\bibnamefont
  {Buitenhuis}}, \bibinfo {author} {\bibfnamefont {M.}~\bibnamefont {Kagias}},
  \bibinfo {author} {\bibfnamefont {K.}~\bibnamefont {Jefimovs}}, \bibinfo
  {author} {\bibfnamefont {F.}~\bibnamefont {Zontone}},\ and\ \bibinfo {author}
  {\bibfnamefont {Y.}~\bibnamefont {Chushkin}},\ }\bibfield  {title} {\bibinfo
  {title} {Anisotropic de {gennes} {narrowing} in {confined} {fluids}},\ }\href
  {https://doi.org/10.1103/PhysRevLett.116.167801} {\bibfield  {journal}
  {\bibinfo  {journal} {Phys. Rev. Lett.}\ }\textbf {\bibinfo {volume} {116}},\
  \bibinfo {pages} {167801} (\bibinfo {year} {2016}{\natexlab{a}})}\BibitemShut
  {NoStop}%
\bibitem [{\citenamefont {Kienle}\ and\ \citenamefont
  {Kuhl}(2016)}]{PhysRevLett.117.036101}%
  \BibitemOpen
  \bibfield  {author} {\bibinfo {author} {\bibfnamefont {D.~F.}\ \bibnamefont
  {Kienle}}\ and\ \bibinfo {author} {\bibfnamefont {T.~L.}\ \bibnamefont
  {Kuhl}},\ }\bibfield  {title} {\bibinfo {title} {Density and {phase} {state}
  of a {confined} {nonpolar} {fluid}},\ }\href
  {https://doi.org/10.1103/PhysRevLett.117.036101} {\bibfield  {journal}
  {\bibinfo  {journal} {Phys. Rev. Lett.}\ }\textbf {\bibinfo {volume} {117}},\
  \bibinfo {pages} {036101} (\bibinfo {year} {2016})}\BibitemShut {NoStop}%
\bibitem [{\citenamefont {Zhang}\ and\ \citenamefont
  {Cheng}(2016)}]{PhysRevLett.116.098302}%
  \BibitemOpen
  \bibfield  {author} {\bibinfo {author} {\bibfnamefont {B.}~\bibnamefont
  {Zhang}}\ and\ \bibinfo {author} {\bibfnamefont {X.}~\bibnamefont {Cheng}},\
  }\bibfield  {title} {\bibinfo {title} {Structures and {dynamics} of
  {glass}-{forming} {colloidal} {liquids} under {spherical} {confinement}},\
  }\href {https://doi.org/10.1103/PhysRevLett.116.098302} {\bibfield  {journal}
  {\bibinfo  {journal} {Phys. Rev. Lett.}\ }\textbf {\bibinfo {volume} {116}},\
  \bibinfo {pages} {098302} (\bibinfo {year} {2016})}\BibitemShut {NoStop}%
\bibitem [{\citenamefont {Nyg\aa{}rd}\ \emph
  {et~al.}(2016{\natexlab{b}})\citenamefont {Nyg\aa{}rd}, \citenamefont
  {Sarman}, \citenamefont {Hyltegren}, \citenamefont {Chodankar}, \citenamefont
  {Perret}, \citenamefont {Buitenhuis}, \citenamefont {van~der Veen},\ and\
  \citenamefont {Kjellander}}]{PhysRevX.6.011014}%
  \BibitemOpen
  \bibfield  {author} {\bibinfo {author} {\bibfnamefont {K.}~\bibnamefont
  {Nyg\aa{}rd}}, \bibinfo {author} {\bibfnamefont {S.}~\bibnamefont {Sarman}},
  \bibinfo {author} {\bibfnamefont {K.}~\bibnamefont {Hyltegren}}, \bibinfo
  {author} {\bibfnamefont {S.}~\bibnamefont {Chodankar}}, \bibinfo {author}
  {\bibfnamefont {E.}~\bibnamefont {Perret}}, \bibinfo {author} {\bibfnamefont
  {J.}~\bibnamefont {Buitenhuis}}, \bibinfo {author} {\bibfnamefont {J.~F.}\
  \bibnamefont {van~der Veen}},\ and\ \bibinfo {author} {\bibfnamefont
  {R.}~\bibnamefont {Kjellander}},\ }\bibfield  {title} {\bibinfo {title}
  {Density {fluctuations} of {hard}-{sphere} {fluids} in {narrow}
  {confinement}},\ }\href {https://doi.org/10.1103/PhysRevX.6.011014}
  {\bibfield  {journal} {\bibinfo  {journal} {Phys. Rev. X}\ }\textbf {\bibinfo
  {volume} {6}},\ \bibinfo {pages} {011014} (\bibinfo {year}
  {2016}{\natexlab{b}})}\BibitemShut {NoStop}%
\bibitem [{\citenamefont {Villada-Balbuena}\ \emph {et~al.}(2022)\citenamefont
  {Villada-Balbuena}, \citenamefont {Jung}, \citenamefont {Zuccolotto-Bernez},
  \citenamefont {Franosch},\ and\ \citenamefont
  {Egelhaaf}}]{villada2022layering}%
  \BibitemOpen
  \bibfield  {author} {\bibinfo {author} {\bibfnamefont {A.}~\bibnamefont
  {Villada-Balbuena}}, \bibinfo {author} {\bibfnamefont {G.}~\bibnamefont
  {Jung}}, \bibinfo {author} {\bibfnamefont {A.~B.}\ \bibnamefont
  {Zuccolotto-Bernez}}, \bibinfo {author} {\bibfnamefont {T.}~\bibnamefont
  {Franosch}},\ and\ \bibinfo {author} {\bibfnamefont {S.~U.}\ \bibnamefont
  {Egelhaaf}},\ }\bibfield  {title} {\bibinfo {title} {Layering and packing in
  confined colloidal suspensions},\ }\href {https://doi.org/10.1039/D2SM00412G}
  {\bibfield  {journal} {\bibinfo  {journal} {Soft Matter}\ }\textbf {\bibinfo
  {volume} {18}},\ \bibinfo {pages} {4699} (\bibinfo {year}
  {2022})}\BibitemShut {NoStop}%
\bibitem [{\citenamefont {Scheidler}\ \emph {et~al.}(2000)\citenamefont
  {Scheidler}, \citenamefont {Kob},\ and\ \citenamefont
  {Binder}}]{Scheidler_2000}%
  \BibitemOpen
  \bibfield  {author} {\bibinfo {author} {\bibfnamefont {P.}~\bibnamefont
  {Scheidler}}, \bibinfo {author} {\bibfnamefont {W.}~\bibnamefont {Kob}},\
  and\ \bibinfo {author} {\bibfnamefont {K.}~\bibnamefont {Binder}},\
  }\bibfield  {title} {\bibinfo {title} {The relaxation dynamics of a simple
  glass former confined in a pore},\ }\href
  {https://doi.org/10.1209/epl/i2000-00435-1} {\bibfield  {journal} {\bibinfo
  {journal} {EPL}\ }\textbf {\bibinfo {volume} {52}},\ \bibinfo {pages} {277}
  (\bibinfo {year} {2000})}\BibitemShut {NoStop}%
\bibitem [{\citenamefont {Scheidler}\ \emph {et~al.}(2002)\citenamefont
  {Scheidler}, \citenamefont {Kob},\ and\ \citenamefont
  {Binder}}]{Scheidler_2002}%
  \BibitemOpen
  \bibfield  {author} {\bibinfo {author} {\bibfnamefont {P.}~\bibnamefont
  {Scheidler}}, \bibinfo {author} {\bibfnamefont {W.}~\bibnamefont {Kob}},\
  and\ \bibinfo {author} {\bibfnamefont {K.}~\bibnamefont {Binder}},\
  }\bibfield  {title} {\bibinfo {title} {Cooperative motion and growing length
  scales in supercooled confined liquids},\ }\href
  {https://doi.org/10.1209/epl/i2002-00182-9} {\bibfield  {journal} {\bibinfo
  {journal} {EPL}\ }\textbf {\bibinfo {volume} {59}},\ \bibinfo {pages} {701}
  (\bibinfo {year} {2002})}\BibitemShut {NoStop}%
\bibitem [{\citenamefont {Scheidler}\ \emph {et~al.}(2004)\citenamefont
  {Scheidler}, \citenamefont {Kob},\ and\ \citenamefont
  {Binder}}]{doi:10.1021/jp036593s}%
  \BibitemOpen
  \bibfield  {author} {\bibinfo {author} {\bibfnamefont {P.}~\bibnamefont
  {Scheidler}}, \bibinfo {author} {\bibfnamefont {W.}~\bibnamefont {Kob}},\
  and\ \bibinfo {author} {\bibfnamefont {K.}~\bibnamefont {Binder}},\
  }\bibfield  {title} {\bibinfo {title} {The {relaxation} {dynamics} of a
  {supercooled} {liquid} {confined} by {rough} {walls}},\ }\href
  {https://doi.org/10.1021/jp036593s} {\bibfield  {journal} {\bibinfo
  {journal} {J. Phys. Chem. B}\ }\textbf {\bibinfo {volume} {108}},\ \bibinfo
  {pages} {6673} (\bibinfo {year} {2004})}\BibitemShut {NoStop}%
\bibitem [{\citenamefont {Varnik}\ \emph {et~al.}(2002)\citenamefont {Varnik},
  \citenamefont {Baschnagel},\ and\ \citenamefont
  {Binder}}]{PhysRevE.65.021507}%
  \BibitemOpen
  \bibfield  {author} {\bibinfo {author} {\bibfnamefont {F.}~\bibnamefont
  {Varnik}}, \bibinfo {author} {\bibfnamefont {J.}~\bibnamefont {Baschnagel}},\
  and\ \bibinfo {author} {\bibfnamefont {K.}~\bibnamefont {Binder}},\
  }\bibfield  {title} {\bibinfo {title} {Reduction of the glass transition
  temperature in polymer films: A molecular-dynamics study},\ }\href
  {https://doi.org/10.1103/PhysRevE.65.021507} {\bibfield  {journal} {\bibinfo
  {journal} {Phys. Rev. E}\ }\textbf {\bibinfo {volume} {65}},\ \bibinfo
  {pages} {021507} (\bibinfo {year} {2002})}\BibitemShut {NoStop}%
\bibitem [{\citenamefont {Torres}\ \emph {et~al.}(2000)\citenamefont {Torres},
  \citenamefont {Nealey},\ and\ \citenamefont
  {de~Pablo}}]{PhysRevLett.85.3221}%
  \BibitemOpen
  \bibfield  {author} {\bibinfo {author} {\bibfnamefont {J.~A.}\ \bibnamefont
  {Torres}}, \bibinfo {author} {\bibfnamefont {P.~F.}\ \bibnamefont {Nealey}},\
  and\ \bibinfo {author} {\bibfnamefont {J.~J.}\ \bibnamefont {de~Pablo}},\
  }\bibfield  {title} {\bibinfo {title} {Molecular {Ssimulation} of {ultrathin}
  {polymeric} {films} near the {glass} {transition}},\ }\href
  {https://doi.org/10.1103/PhysRevLett.85.3221} {\bibfield  {journal} {\bibinfo
   {journal} {Phys. Rev. Lett.}\ }\textbf {\bibinfo {volume} {85}},\ \bibinfo
  {pages} {3221} (\bibinfo {year} {2000})}\BibitemShut {NoStop}%
\bibitem [{\citenamefont {Baschnagel}\ and\ \citenamefont
  {Varnik}(2005)}]{Baschnagel_2005}%
  \BibitemOpen
  \bibfield  {author} {\bibinfo {author} {\bibfnamefont {J.}~\bibnamefont
  {Baschnagel}}\ and\ \bibinfo {author} {\bibfnamefont {F.}~\bibnamefont
  {Varnik}},\ }\bibfield  {title} {\bibinfo {title} {Computer simulations of
  supercooled polymer melts in the bulk and in confined geometry},\ }\href
  {https://doi.org/10.1088/0953-8984/17/32/r02} {\bibfield  {journal} {\bibinfo
   {journal} {J. Phys. Condens. Matter}\ }\textbf {\bibinfo {volume} {17}},\
  \bibinfo {pages} {R851} (\bibinfo {year} {2005})}\BibitemShut {NoStop}%
\bibitem [{\citenamefont {Mittal}\ \emph {et~al.}(2006)\citenamefont {Mittal},
  \citenamefont {Errington},\ and\ \citenamefont
  {Truskett}}]{PhysRevLett.96.177804}%
  \BibitemOpen
  \bibfield  {author} {\bibinfo {author} {\bibfnamefont {J.}~\bibnamefont
  {Mittal}}, \bibinfo {author} {\bibfnamefont {J.~R.}\ \bibnamefont
  {Errington}},\ and\ \bibinfo {author} {\bibfnamefont {T.~M.}\ \bibnamefont
  {Truskett}},\ }\bibfield  {title} {\bibinfo {title} {Thermodynamics
  {predicts} {how} {confinement} {modifies} the {dynamics} of the {equilibrium}
  {hard}-{sphere} {fluid}},\ }\href
  {https://doi.org/10.1103/PhysRevLett.96.177804} {\bibfield  {journal}
  {\bibinfo  {journal} {Phys. Rev. Lett.}\ }\textbf {\bibinfo {volume} {96}},\
  \bibinfo {pages} {177804} (\bibinfo {year} {2006})}\BibitemShut {NoStop}%
\bibitem [{\citenamefont {Mittal}\ \emph
  {et~al.}(2007{\natexlab{a}})\citenamefont {Mittal}, \citenamefont {Shen},
  \citenamefont {Errington},\ and\ \citenamefont
  {Truskett}}]{doi:10.1063/1.2795699}%
  \BibitemOpen
  \bibfield  {author} {\bibinfo {author} {\bibfnamefont {J.}~\bibnamefont
  {Mittal}}, \bibinfo {author} {\bibfnamefont {V.~K.}\ \bibnamefont {Shen}},
  \bibinfo {author} {\bibfnamefont {J.~R.}\ \bibnamefont {Errington}},\ and\
  \bibinfo {author} {\bibfnamefont {T.~M.}\ \bibnamefont {Truskett}},\
  }\bibfield  {title} {\bibinfo {title} {Confinement, entropy, and
  single-particle dynamics of equilibrium hard-sphere mixtures},\ }\href
  {https://doi.org/10.1063/1.2795699} {\bibfield  {journal} {\bibinfo
  {journal} {J. Chem. Phys.}\ }\textbf {\bibinfo {volume} {127}},\ \bibinfo
  {pages} {154513} (\bibinfo {year} {2007}{\natexlab{a}})}\BibitemShut
  {NoStop}%
\bibitem [{\citenamefont {Mittal}\ \emph
  {et~al.}(2007{\natexlab{b}})\citenamefont {Mittal}, \citenamefont
  {Errington},\ and\ \citenamefont {Truskett}}]{doi:10.1021/jp071369e}%
  \BibitemOpen
  \bibfield  {author} {\bibinfo {author} {\bibfnamefont {J.}~\bibnamefont
  {Mittal}}, \bibinfo {author} {\bibfnamefont {J.~R.}\ \bibnamefont
  {Errington}},\ and\ \bibinfo {author} {\bibfnamefont {T.~M.}\ \bibnamefont
  {Truskett}},\ }\bibfield  {title} {\bibinfo {title} {Relationships between
  {self}-{diffusivity}, {packing} {fraction}, and {excess} {entropy} in
  {simple} {bulk} and {confined} {fluids}},\ }\href
  {https://doi.org/10.1021/jp071369e} {\bibfield  {journal} {\bibinfo
  {journal} {J. Phys. Chem. B}\ }\textbf {\bibinfo {volume} {111}},\ \bibinfo
  {pages} {10054} (\bibinfo {year} {2007}{\natexlab{b}})}\BibitemShut {NoStop}%
\bibitem [{\citenamefont {Saw}\ and\ \citenamefont
  {Dasgupta}(2016)}]{doi:10.1063/1.4959942}%
  \BibitemOpen
  \bibfield  {author} {\bibinfo {author} {\bibfnamefont {S.}~\bibnamefont
  {Saw}}\ and\ \bibinfo {author} {\bibfnamefont {C.}~\bibnamefont {Dasgupta}},\
  }\bibfield  {title} {\bibinfo {title} {Role of density modulation in the
  spatially resolved dynamics of strongly confined liquids},\ }\href
  {https://doi.org/10.1063/1.4959942} {\bibfield  {journal} {\bibinfo
  {journal} {J. Chem. Phys.}\ }\textbf {\bibinfo {volume} {145}},\ \bibinfo
  {pages} {054707} (\bibinfo {year} {2016})}\BibitemShut {NoStop}%
\bibitem [{\citenamefont {Goel}\ \emph {et~al.}(2008)\citenamefont {Goel},
  \citenamefont {Krekelberg}, \citenamefont {Errington},\ and\ \citenamefont
  {Truskett}}]{PhysRevLett.100.106001}%
  \BibitemOpen
  \bibfield  {author} {\bibinfo {author} {\bibfnamefont {G.}~\bibnamefont
  {Goel}}, \bibinfo {author} {\bibfnamefont {W.~P.}\ \bibnamefont
  {Krekelberg}}, \bibinfo {author} {\bibfnamefont {J.~R.}\ \bibnamefont
  {Errington}},\ and\ \bibinfo {author} {\bibfnamefont {T.~M.}\ \bibnamefont
  {Truskett}},\ }\bibfield  {title} {\bibinfo {title} {Tuning {density}
  {profiles} and {mobility} of {inhomogeneous} {fluids}},\ }\href
  {https://doi.org/10.1103/PhysRevLett.100.106001} {\bibfield  {journal}
  {\bibinfo  {journal} {Phys. Rev. Lett.}\ }\textbf {\bibinfo {volume} {100}},\
  \bibinfo {pages} {106001} (\bibinfo {year} {2008})}\BibitemShut {NoStop}%
\bibitem [{\citenamefont {Krekelberg}\ \emph {et~al.}(2011)\citenamefont
  {Krekelberg}, \citenamefont {Shen}, \citenamefont {Errington},\ and\
  \citenamefont {Truskett}}]{doi:10.1063/1.3651478}%
  \BibitemOpen
  \bibfield  {author} {\bibinfo {author} {\bibfnamefont {W.~P.}\ \bibnamefont
  {Krekelberg}}, \bibinfo {author} {\bibfnamefont {V.~K.}\ \bibnamefont
  {Shen}}, \bibinfo {author} {\bibfnamefont {J.~R.}\ \bibnamefont
  {Errington}},\ and\ \bibinfo {author} {\bibfnamefont {T.~M.}\ \bibnamefont
  {Truskett}},\ }\bibfield  {title} {\bibinfo {title} {Impact of surface
  roughness on diffusion of confined fluids},\ }\href
  {https://doi.org/10.1063/1.3651478} {\bibfield  {journal} {\bibinfo
  {journal} {J. Chem. Phys.}\ }\textbf {\bibinfo {volume} {135}},\ \bibinfo
  {pages} {154502} (\bibinfo {year} {2011})}\BibitemShut {NoStop}%
\bibitem [{\citenamefont {Ingebrigtsen}\ \emph {et~al.}(2013)\citenamefont
  {Ingebrigtsen}, \citenamefont {Errington}, \citenamefont {Truskett},\ and\
  \citenamefont {Dyre}}]{PhysRevLett.111.235901}%
  \BibitemOpen
  \bibfield  {author} {\bibinfo {author} {\bibfnamefont {T.~S.}\ \bibnamefont
  {Ingebrigtsen}}, \bibinfo {author} {\bibfnamefont {J.~R.}\ \bibnamefont
  {Errington}}, \bibinfo {author} {\bibfnamefont {T.~M.}\ \bibnamefont
  {Truskett}},\ and\ \bibinfo {author} {\bibfnamefont {J.~C.}\ \bibnamefont
  {Dyre}},\ }\bibfield  {title} {\bibinfo {title} {Predicting {how}
  {nanoconfinement} {changes} the {relaxation} {time} of a {supercooled}
  {liquid}},\ }\href {https://doi.org/10.1103/PhysRevLett.111.235901}
  {\bibfield  {journal} {\bibinfo  {journal} {Phys. Rev. Lett.}\ }\textbf
  {\bibinfo {volume} {111}},\ \bibinfo {pages} {235901} (\bibinfo {year}
  {2013})}\BibitemShut {NoStop}%
\bibitem [{\citenamefont {Krishnan}\ and\ \citenamefont
  {Ayappa}(2003)}]{doi:10.1063/1.1524191}%
  \BibitemOpen
  \bibfield  {author} {\bibinfo {author} {\bibfnamefont {S.~H.}\ \bibnamefont
  {Krishnan}}\ and\ \bibinfo {author} {\bibfnamefont {K.~G.}\ \bibnamefont
  {Ayappa}},\ }\bibfield  {title} {\bibinfo {title} {Modeling velocity
  autocorrelation functions of confined fluids: A memory function approach},\
  }\href {https://doi.org/10.1063/1.1524191} {\bibfield  {journal} {\bibinfo
  {journal} {J. Chem. Phys.}\ }\textbf {\bibinfo {volume} {118}},\ \bibinfo
  {pages} {690} (\bibinfo {year} {2003})}\BibitemShut {NoStop}%
\bibitem [{\citenamefont {Krishnan}\ and\ \citenamefont
  {Ayappa}(2012)}]{PhysRevE.86.011504}%
  \BibitemOpen
  \bibfield  {author} {\bibinfo {author} {\bibfnamefont {S.~H.}\ \bibnamefont
  {Krishnan}}\ and\ \bibinfo {author} {\bibfnamefont {K.~G.}\ \bibnamefont
  {Ayappa}},\ }\bibfield  {title} {\bibinfo {title} {Glassy dynamics in a
  confined monatomic fluid},\ }\href
  {https://doi.org/10.1103/PhysRevE.86.011504} {\bibfield  {journal} {\bibinfo
  {journal} {Phys. Rev. E}\ }\textbf {\bibinfo {volume} {86}},\ \bibinfo
  {pages} {011504} (\bibinfo {year} {2012})}\BibitemShut {NoStop}%
\bibitem [{\citenamefont {Ingebrigtsen}\ and\ \citenamefont
  {Dyre}(2014)}]{C3SM52441H}%
  \BibitemOpen
  \bibfield  {author} {\bibinfo {author} {\bibfnamefont {T.~S.}\ \bibnamefont
  {Ingebrigtsen}}\ and\ \bibinfo {author} {\bibfnamefont {J.~C.}\ \bibnamefont
  {Dyre}},\ }\bibfield  {title} {\bibinfo {title} {The impact range for smooth
  wall–liquid interactions in nanoconfined liquids},\ }\href
  {https://doi.org/10.1039/C3SM52441H} {\bibfield  {journal} {\bibinfo
  {journal} {Soft Matter}\ }\textbf {\bibinfo {volume} {10}},\ \bibinfo {pages}
  {4324} (\bibinfo {year} {2014})}\BibitemShut {NoStop}%
\bibitem [{\citenamefont {Mandal}\ \emph {et~al.}(2017)\citenamefont {Mandal},
  \citenamefont {Lang}, \citenamefont {Bo\ifmmode~\mbox{\c{t}}\else
  \c{t}\fi{}an},\ and\ \citenamefont {Franosch}}]{Mandal2017a}%
  \BibitemOpen
  \bibfield  {author} {\bibinfo {author} {\bibfnamefont {S.}~\bibnamefont
  {Mandal}}, \bibinfo {author} {\bibfnamefont {S.}~\bibnamefont {Lang}},
  \bibinfo {author} {\bibfnamefont {V.}~\bibnamefont
  {Bo\ifmmode~\mbox{\c{t}}\else \c{t}\fi{}an}},\ and\ \bibinfo {author}
  {\bibfnamefont {T.}~\bibnamefont {Franosch}},\ }\bibfield  {title} {\bibinfo
  {title} {Nonergodicity parameters of confined hard-sphere glasses},\ }\href
  {https://doi.org/10.1039/C7SM00905D} {\bibfield  {journal} {\bibinfo
  {journal} {Soft Matter}\ }\textbf {\bibinfo {volume} {13}},\ \bibinfo {pages}
  {6167} (\bibinfo {year} {2017})}\BibitemShut {NoStop}%
\bibitem [{\citenamefont {Jung}\ and\ \citenamefont
  {Franosch}(2022)}]{Jung2022_extreme}%
  \BibitemOpen
  \bibfield  {author} {\bibinfo {author} {\bibfnamefont {G.}~\bibnamefont
  {Jung}}\ and\ \bibinfo {author} {\bibfnamefont {T.}~\bibnamefont
  {Franosch}},\ }\bibfield  {title} {\bibinfo {title} {Structural properties of
  liquids in extreme confinement},\ }\href
  {https://doi.org/10.1103/PhysRevE.106.014614} {\bibfield  {journal} {\bibinfo
   {journal} {Phys. Rev. E}\ }\textbf {\bibinfo {volume} {106}},\ \bibinfo
  {pages} {014614} (\bibinfo {year} {2022})}\BibitemShut {NoStop}%
\bibitem [{\citenamefont {Lang}\ \emph {et~al.}(2012)\citenamefont {Lang},
  \citenamefont {Schilling}, \citenamefont {Krakoviack},\ and\ \citenamefont
  {Franosch}}]{Lang2012}%
  \BibitemOpen
  \bibfield  {author} {\bibinfo {author} {\bibfnamefont {S.}~\bibnamefont
  {Lang}}, \bibinfo {author} {\bibfnamefont {R.}~\bibnamefont {Schilling}},
  \bibinfo {author} {\bibfnamefont {V.}~\bibnamefont {Krakoviack}},\ and\
  \bibinfo {author} {\bibfnamefont {T.}~\bibnamefont {Franosch}},\ }\bibfield
  {title} {\bibinfo {title} {Mode-coupling theory of the glass transition for
  confined fluids},\ }\href {https://doi.org/10.1103/PhysRevE.86.021502}
  {\bibfield  {journal} {\bibinfo  {journal} {Phys. Rev. E}\ }\textbf {\bibinfo
  {volume} {86}},\ \bibinfo {pages} {021502} (\bibinfo {year}
  {2012})}\BibitemShut {NoStop}%
\bibitem [{\citenamefont {Jung}\ \emph
  {et~al.}(2020{\natexlab{a}})\citenamefont {Jung}, \citenamefont {Caraglio},
  \citenamefont {Schrack},\ and\ \citenamefont {Franosch}}]{Jung:2020}%
  \BibitemOpen
  \bibfield  {author} {\bibinfo {author} {\bibfnamefont {G.}~\bibnamefont
  {Jung}}, \bibinfo {author} {\bibfnamefont {M.}~\bibnamefont {Caraglio}},
  \bibinfo {author} {\bibfnamefont {L.}~\bibnamefont {Schrack}},\ and\ \bibinfo
  {author} {\bibfnamefont {T.}~\bibnamefont {Franosch}},\ }\bibfield  {title}
  {\bibinfo {title} {Dynamical properties of densely packed confined
  hard-sphere fluids},\ }\href {https://doi.org/10.1103/PhysRevE.102.012612}
  {\bibfield  {journal} {\bibinfo  {journal} {Phys. Rev. E}\ }\textbf {\bibinfo
  {volume} {102}},\ \bibinfo {pages} {012612} (\bibinfo {year}
  {2020}{\natexlab{a}})}\BibitemShut {NoStop}%
\bibitem [{\citenamefont {Jung}\ and\ \citenamefont
  {Petersen}(2020)}]{Jung2020_cryst}%
  \BibitemOpen
  \bibfield  {author} {\bibinfo {author} {\bibfnamefont {G.}~\bibnamefont
  {Jung}}\ and\ \bibinfo {author} {\bibfnamefont {C.~F.}\ \bibnamefont
  {Petersen}},\ }\bibfield  {title} {\bibinfo {title} {Confinement-induced
  demixing and crystallization},\ }\href
  {https://doi.org/10.1103/PhysRevResearch.2.033207} {\bibfield  {journal}
  {\bibinfo  {journal} {Phys. Rev. Research}\ }\textbf {\bibinfo {volume}
  {2}},\ \bibinfo {pages} {033207} (\bibinfo {year} {2020})}\BibitemShut
  {NoStop}%
\bibitem [{\citenamefont {Jung}\ \emph
  {et~al.}(2020{\natexlab{b}})\citenamefont {Jung}, \citenamefont {Schrack},\
  and\ \citenamefont {Franosch}}]{Jung2020_self}%
  \BibitemOpen
  \bibfield  {author} {\bibinfo {author} {\bibfnamefont {G.}~\bibnamefont
  {Jung}}, \bibinfo {author} {\bibfnamefont {L.}~\bibnamefont {Schrack}},\ and\
  \bibinfo {author} {\bibfnamefont {T.}~\bibnamefont {Franosch}},\ }\bibfield
  {title} {\bibinfo {title} {Tagged-particle dynamics in confined colloidal
  liquids},\ }\href {https://doi.org/10.1103/PhysRevE.102.032611} {\bibfield
  {journal} {\bibinfo  {journal} {Phys. Rev. E}\ }\textbf {\bibinfo {volume}
  {102}},\ \bibinfo {pages} {032611} (\bibinfo {year}
  {2020}{\natexlab{b}})}\BibitemShut {NoStop}%
\bibitem [{\citenamefont {Yamchi}\ \emph {et~al.}(2012)\citenamefont {Yamchi},
  \citenamefont {Ashwin},\ and\ \citenamefont {Bowles}}]{Yamchi2012_disks}%
  \BibitemOpen
  \bibfield  {author} {\bibinfo {author} {\bibfnamefont {M.~Z.}\ \bibnamefont
  {Yamchi}}, \bibinfo {author} {\bibfnamefont {S.~S.}\ \bibnamefont {Ashwin}},\
  and\ \bibinfo {author} {\bibfnamefont {R.~K.}\ \bibnamefont {Bowles}},\
  }\bibfield  {title} {\bibinfo {title} {Fragile-strong fluid crossover and
  universal relaxation times in a confined hard-disk fluid},\ }\href
  {https://doi.org/10.1103/PhysRevLett.109.225701} {\bibfield  {journal}
  {\bibinfo  {journal} {Phys. Rev. Lett.}\ }\textbf {\bibinfo {volume} {109}},\
  \bibinfo {pages} {225701} (\bibinfo {year} {2012})}\BibitemShut {NoStop}%
\bibitem [{\citenamefont {Lang}\ \emph {et~al.}(2010)\citenamefont {Lang},
  \citenamefont {Botan}, \citenamefont {Oettel}, \citenamefont {Hajnal},
  \citenamefont {Franosch},\ and\ \citenamefont {Schilling}}]{Lang2010}%
  \BibitemOpen
  \bibfield  {author} {\bibinfo {author} {\bibfnamefont {S.}~\bibnamefont
  {Lang}}, \bibinfo {author} {\bibfnamefont {V.}~\bibnamefont {Botan}},
  \bibinfo {author} {\bibfnamefont {M.}~\bibnamefont {Oettel}}, \bibinfo
  {author} {\bibfnamefont {D.}~\bibnamefont {Hajnal}}, \bibinfo {author}
  {\bibfnamefont {T.}~\bibnamefont {Franosch}},\ and\ \bibinfo {author}
  {\bibfnamefont {R.}~\bibnamefont {Schilling}},\ }\bibfield  {title} {\bibinfo
  {title} {Glass {Transition} in {Confined} {Geometry}},\ }\href
  {https://doi.org/10.1103/PhysRevLett.105.125701} {\bibfield  {journal}
  {\bibinfo  {journal} {Phys. Rev. Lett.}\ }\textbf {\bibinfo {volume} {105}},\
  \bibinfo {pages} {125701} (\bibinfo {year} {2010})}\BibitemShut {NoStop}%
\bibitem [{\citenamefont {Lang}\ \emph {et~al.}(2013)\citenamefont {Lang},
  \citenamefont {Schilling},\ and\ \citenamefont {Franosch}}]{Lang_2013}%
  \BibitemOpen
  \bibfield  {author} {\bibinfo {author} {\bibfnamefont {S.}~\bibnamefont
  {Lang}}, \bibinfo {author} {\bibfnamefont {R.}~\bibnamefont {Schilling}},\
  and\ \bibinfo {author} {\bibfnamefont {T.}~\bibnamefont {Franosch}},\
  }\bibfield  {title} {\bibinfo {title} {Mode-coupling theory for multiple
  decay channels},\ }\href {https://doi.org/10.1088/1742-5468/2013/12/P12007}
  {\bibfield  {journal} {\bibinfo  {journal} {Journal of Statistical Mechanics:
  Theory and Experiment}\ }\textbf {\bibinfo {volume} {2013}},\ \bibinfo
  {pages} {P12007} (\bibinfo {year} {2013})}\BibitemShut {NoStop}%
\bibitem [{\citenamefont {Bengtzelius}\ \emph {et~al.}(1984)\citenamefont
  {Bengtzelius}, \citenamefont {G{\"o}tze},\ and\ \citenamefont
  {Sj{\"o}lander}}]{Bengtzelius_1984}%
  \BibitemOpen
  \bibfield  {author} {\bibinfo {author} {\bibfnamefont {U.}~\bibnamefont
  {Bengtzelius}}, \bibinfo {author} {\bibfnamefont {W.}~\bibnamefont
  {G{\"o}tze}},\ and\ \bibinfo {author} {\bibfnamefont {A.}~\bibnamefont
  {Sj{\"o}lander}},\ }\bibfield  {title} {\bibinfo {title} {Dynamics of
  supercooled liquids and the glass transition},\ }\href
  {https://doi.org/10.1088/0022-3719/17/33/005} {\bibfield  {journal} {\bibinfo
   {journal} {J. Phys. C: Solid State Phys.}\ }\textbf {\bibinfo {volume}
  {17}},\ \bibinfo {pages} {5915} (\bibinfo {year} {1984})}\BibitemShut
  {NoStop}%
\bibitem [{\citenamefont {G{\"{o}}tze}(2009)}]{Gotze2009}%
  \BibitemOpen
  \bibfield  {author} {\bibinfo {author} {\bibfnamefont {W.}~\bibnamefont
  {G{\"{o}}tze}},\ }\href@noop {} {\emph {\bibinfo {title} {{Complex Dynamics
  of Glass-Forming Liquids - A Mode-Coupling Theory}}}}\ (\bibinfo  {publisher}
  {Oxford University Press},\ \bibinfo {address} {Oxford},\ \bibinfo {year}
  {2009})\BibitemShut {NoStop}%
\bibitem [{\citenamefont {Janssen}(2018)}]{JansenReview2018}%
  \BibitemOpen
  \bibfield  {author} {\bibinfo {author} {\bibfnamefont {L.~M.~C.}\
  \bibnamefont {Janssen}},\ }\bibfield  {title} {\bibinfo {title}
  {{Mode-Coupling Theory of the Glass Transition: A Primer}},\ }\href
  {https://doi.org/10.3389/fphy.2018.00097} {\bibfield  {journal} {\bibinfo
  {journal} {Frontiers in Physics}\ }\textbf {\bibinfo {volume} {6}},\ \bibinfo
  {pages} {97} (\bibinfo {year} {2018})}\BibitemShut {NoStop}%
\bibitem [{\citenamefont {Alder}\ and\ \citenamefont
  {Wainwright}(1957{\natexlab{a}})}]{Alder1957a}%
  \BibitemOpen
  \bibfield  {author} {\bibinfo {author} {\bibfnamefont {B.~J.}\ \bibnamefont
  {Alder}}\ and\ \bibinfo {author} {\bibfnamefont {T.~E.}\ \bibnamefont
  {Wainwright}},\ }\bibfield  {title} {\bibinfo {title} {Phase {Transition} for
  a {Hard} {Sphere} {System}},\ }\href {https://doi.org/10.1063/1.1743957}
  {\bibfield  {journal} {\bibinfo  {journal} {J. Chem. Phys.}\ }\textbf
  {\bibinfo {volume} {27}},\ \bibinfo {pages} {1208} (\bibinfo {year}
  {1957}{\natexlab{a}})}\BibitemShut {NoStop}%
\bibitem [{\citenamefont {Rapaport}(1980{\natexlab{a}})}]{Rapaport1980}%
  \BibitemOpen
  \bibfield  {author} {\bibinfo {author} {\bibfnamefont {D.}~\bibnamefont
  {Rapaport}},\ }\bibfield  {title} {\bibinfo {title} {The event scheduling
  problem in molecular dynamic simulation},\ }\href
  {https://doi.org/https://doi.org/10.1016/0021-9991(80)90104-7} {\bibfield
  {journal} {\bibinfo  {journal} {J. Comput. Phys.}\ }\textbf {\bibinfo
  {volume} {34}},\ \bibinfo {pages} {184 } (\bibinfo {year}
  {1980}{\natexlab{a}})}\BibitemShut {NoStop}%
\bibitem [{\citenamefont {Bannerman}\ \emph
  {et~al.}(2011{\natexlab{a}})\citenamefont {Bannerman}, \citenamefont
  {Sargant},\ and\ \citenamefont {Lue}}]{Bannerman2011}%
  \BibitemOpen
  \bibfield  {author} {\bibinfo {author} {\bibfnamefont {M.~N.}\ \bibnamefont
  {Bannerman}}, \bibinfo {author} {\bibfnamefont {R.}~\bibnamefont {Sargant}},\
  and\ \bibinfo {author} {\bibfnamefont {L.}~\bibnamefont {Lue}},\ }\bibfield
  {title} {\bibinfo {title} {Dynam{O}: a free $\mathcal{O}$({N}) general
  event-driven molecular dynamics simulator},\ }\href
  {https://doi.org/10.1002/jcc.21915} {\bibfield  {journal} {\bibinfo
  {journal} {J. Comput. Chem.}\ }\textbf {\bibinfo {volume} {32}},\ \bibinfo
  {pages} {3329} (\bibinfo {year} {2011}{\natexlab{a}})}\BibitemShut {NoStop}%
\bibitem [{\citenamefont {Schrack}\ and\ \citenamefont
  {Franosch}(2020)}]{Schrack:2020a}%
  \BibitemOpen
  \bibfield  {author} {\bibinfo {author} {\bibfnamefont {L.}~\bibnamefont
  {Schrack}}\ and\ \bibinfo {author} {\bibfnamefont {T.}~\bibnamefont
  {Franosch}},\ }\bibfield  {title} {\bibinfo {title} {Mode-coupling theory of
  the glass transition for colloidal liquids in slit geometry},\ }\href
  {https://doi.org/10.1080/14786435.2020.1722859} {\bibfield  {journal}
  {\bibinfo  {journal} {Philosophical Magazine}\ }\textbf {\bibinfo {volume}
  {100}},\ \bibinfo {pages} {1032} (\bibinfo {year} {2020})},\ \bibinfo {note}
  {pMID: 32308566}\BibitemShut {NoStop}%
\bibitem [{\citenamefont {Dhont}(1996)}]{Dhont1996}%
  \BibitemOpen
  \bibfield  {author} {\bibinfo {author} {\bibfnamefont {J.~K.~G.}\
  \bibnamefont {Dhont}},\ }\href@noop {} {\emph {\bibinfo {title} {{An
  introduction to dynamics of colloids}}}}\ (\bibinfo  {publisher} {Elsevier},\
  \bibinfo {year} {1996})\BibitemShut {NoStop}%
\bibitem [{\citenamefont {Zwanzig}(2001)}]{Zwanzig2001}%
  \BibitemOpen
  \bibfield  {author} {\bibinfo {author} {\bibfnamefont {R.}~\bibnamefont
  {Zwanzig}},\ }\href@noop {} {\emph {\bibinfo {title} {{Nonequilibrium
  statistical mechanics}}}}\ (\bibinfo  {publisher} {Oxford University Press},\
  \bibinfo {year} {2001})\BibitemShut {NoStop}%
\bibitem [{\citenamefont {Hansen}\ and\ \citenamefont
  {McDonald}(2013)}]{Hansen:Theory_of_Simple_Liquids}%
  \BibitemOpen
  \bibfield  {author} {\bibinfo {author} {\bibfnamefont {J.~P.}\ \bibnamefont
  {Hansen}}\ and\ \bibinfo {author} {\bibfnamefont {I.~R.}\ \bibnamefont
  {McDonald}},\ }\href {https://doi.org/10.1016/C2010-0-66723-X} {\emph
  {\bibinfo {title} {Theory of {Simple} {Liquids}}}}\ (\bibinfo  {publisher}
  {Academic Press},\ \bibinfo {year} {2013})\BibitemShut {NoStop}%
\bibitem [{\citenamefont {Lang}\ and\ \citenamefont
  {Franosch}(2014)}]{Lang2014b}%
  \BibitemOpen
  \bibfield  {author} {\bibinfo {author} {\bibfnamefont {S.}~\bibnamefont
  {Lang}}\ and\ \bibinfo {author} {\bibfnamefont {T.}~\bibnamefont
  {Franosch}},\ }\bibfield  {title} {\bibinfo {title} {Tagged-particle motion
  in a dense confined liquid},\ }\href
  {https://doi.org/10.1103/PhysRevE.89.062122} {\bibfield  {journal} {\bibinfo
  {journal} {Phys. Rev. E}\ }\textbf {\bibinfo {volume} {89}},\ \bibinfo
  {pages} {062122} (\bibinfo {year} {2014})}\BibitemShut {NoStop}%
\bibitem [{\citenamefont {Rosenfeld}(1989)}]{fmt:Rosenfeld1989}%
  \BibitemOpen
  \bibfield  {author} {\bibinfo {author} {\bibfnamefont {Y.}~\bibnamefont
  {Rosenfeld}},\ }\bibfield  {title} {\bibinfo {title} {Free-energy model for
  the inhomogeneous hard-sphere fluid mixture and density-functional theory of
  freezing},\ }\href {https://doi.org/10.1103/PhysRevLett.63.980} {\bibfield
  {journal} {\bibinfo  {journal} {Phys. Rev. Lett.}\ }\textbf {\bibinfo
  {volume} {63}},\ \bibinfo {pages} {980} (\bibinfo {year} {1989})}\BibitemShut
  {NoStop}%
\bibitem [{\citenamefont {Roth}(2010)}]{fmt:Roth2010}%
  \BibitemOpen
  \bibfield  {author} {\bibinfo {author} {\bibfnamefont {R.}~\bibnamefont
  {Roth}},\ }\bibfield  {title} {\bibinfo {title} {Fundamental measure theory
  for hard-sphere mixtures: a review},\ }\href
  {https://doi.org/10.1088/0953-8984/22/6/063102} {\bibfield  {journal}
  {\bibinfo  {journal} {Journal of Physics: Condensed Matter}\ }\textbf
  {\bibinfo {volume} {22}},\ \bibinfo {pages} {063102} (\bibinfo {year}
  {2010})}\BibitemShut {NoStop}%
\bibitem [{\citenamefont {Lang}(2010)}]{Lang2010D}%
  \BibitemOpen
  \bibfield  {author} {\bibinfo {author} {\bibfnamefont {S.}~\bibnamefont
  {Lang}},\ }\href@noop {} {\emph {\bibinfo {title} {{Glass transition in
  confined geometries described by mode-coupling theory. Diploma thesis,
  Johannes Gutenberg-Universit{\"{a}}t Mainz}}}}\ (\bibinfo {year}
  {2010})\BibitemShut {NoStop}%
\bibitem [{\citenamefont {Petersen}\ \emph {et~al.}(2019)\citenamefont
  {Petersen}, \citenamefont {Schrack},\ and\ \citenamefont
  {Franosch}}]{Petersen_2019}%
  \BibitemOpen
  \bibfield  {author} {\bibinfo {author} {\bibfnamefont {C.~F.}\ \bibnamefont
  {Petersen}}, \bibinfo {author} {\bibfnamefont {L.}~\bibnamefont {Schrack}},\
  and\ \bibinfo {author} {\bibfnamefont {T.}~\bibnamefont {Franosch}},\
  }\bibfield  {title} {\bibinfo {title} {Static properties of quasi-confined
  hard-sphere fluids},\ }\href {https://doi.org/10.1088/1742-5468/ab3342}
  {\bibfield  {journal} {\bibinfo  {journal} {J. Stat. Mech. Theory Exp.}\
  }\textbf {\bibinfo {volume} {2019}},\ \bibinfo {pages} {083216} (\bibinfo
  {year} {2019})}\BibitemShut {NoStop}%
\bibitem [{\citenamefont {Alder}\ and\ \citenamefont
  {Wainwright}(1957{\natexlab{b}})}]{edmd:alder1957}%
  \BibitemOpen
  \bibfield  {author} {\bibinfo {author} {\bibfnamefont {B.~J.}\ \bibnamefont
  {Alder}}\ and\ \bibinfo {author} {\bibfnamefont {T.~E.}\ \bibnamefont
  {Wainwright}},\ }\bibfield  {title} {\bibinfo {title} {Phase transition for a
  hard sphere system},\ }\href {https://doi.org/10.1063/1.1743957} {\bibfield
  {journal} {\bibinfo  {journal} {The Journal of Chemical Physics}\ }\textbf
  {\bibinfo {volume} {27}},\ \bibinfo {pages} {1208} (\bibinfo {year}
  {1957}{\natexlab{b}})}\BibitemShut {NoStop}%
\bibitem [{\citenamefont {Rapaport}(1980{\natexlab{b}})}]{edmd:RAPAPORT1980}%
  \BibitemOpen
  \bibfield  {author} {\bibinfo {author} {\bibfnamefont {D.}~\bibnamefont
  {Rapaport}},\ }\bibfield  {title} {\bibinfo {title} {The event scheduling
  problem in molecular dynamic simulation},\ }\href
  {https://doi.org/https://doi.org/10.1016/0021-9991(80)90104-7} {\bibfield
  {journal} {\bibinfo  {journal} {Journal of Computational Physics}\ }\textbf
  {\bibinfo {volume} {34}},\ \bibinfo {pages} {184 } (\bibinfo {year}
  {1980}{\natexlab{b}})}\BibitemShut {NoStop}%
\bibitem [{\citenamefont {Bannerman}\ \emph
  {et~al.}(2011{\natexlab{b}})\citenamefont {Bannerman}, \citenamefont
  {Sargant},\ and\ \citenamefont {Lue}}]{edmd:bannerman2011}%
  \BibitemOpen
  \bibfield  {author} {\bibinfo {author} {\bibfnamefont {M.~N.}\ \bibnamefont
  {Bannerman}}, \bibinfo {author} {\bibfnamefont {R.}~\bibnamefont {Sargant}},\
  and\ \bibinfo {author} {\bibfnamefont {L.}~\bibnamefont {Lue}},\ }\bibfield
  {title} {\bibinfo {title} {Dynamo: a free $\mathcal{O}(n)$ general
  event-driven molecular dynamics simulator},\ }\href
  {https://doi.org/10.1002/jcc.21915} {\bibfield  {journal} {\bibinfo
  {journal} {Journal of Computational Chemistry}\ }\textbf {\bibinfo {volume}
  {32}},\ \bibinfo {pages} {3329} (\bibinfo {year}
  {2011}{\natexlab{b}})}\BibitemShut {NoStop}%
\bibitem [{\citenamefont {Gleim}\ \emph {et~al.}(1998)\citenamefont {Gleim},
  \citenamefont {Kob},\ and\ \citenamefont {Binder}}]{PhysRevLett.81.4404}%
  \BibitemOpen
  \bibfield  {author} {\bibinfo {author} {\bibfnamefont {T.}~\bibnamefont
  {Gleim}}, \bibinfo {author} {\bibfnamefont {W.}~\bibnamefont {Kob}},\ and\
  \bibinfo {author} {\bibfnamefont {K.}~\bibnamefont {Binder}},\ }\bibfield
  {title} {\bibinfo {title} {How does the relaxation of a supercooled liquid
  depend on its microscopic dynamics?},\ }\href
  {https://doi.org/10.1103/PhysRevLett.81.4404} {\bibfield  {journal} {\bibinfo
   {journal} {Phys. Rev. Lett.}\ }\textbf {\bibinfo {volume} {81}},\ \bibinfo
  {pages} {4404} (\bibinfo {year} {1998})}\BibitemShut {NoStop}%
\bibitem [{\citenamefont {Kob}\ and\ \citenamefont {Andersen}(1994)}]{Kob1994}%
  \BibitemOpen
  \bibfield  {author} {\bibinfo {author} {\bibfnamefont {W.}~\bibnamefont
  {Kob}}\ and\ \bibinfo {author} {\bibfnamefont {H.~C.}\ \bibnamefont
  {Andersen}},\ }\bibfield  {title} {\bibinfo {title} {Scaling {behavior} in
  the $\ensuremath{\beta}$-{relaxation} {regime} of a {supercooled}
  {Lennard}-{Jones} {mixture}},\ }\href
  {https://doi.org/10.1103/PhysRevLett.73.1376} {\bibfield  {journal} {\bibinfo
   {journal} {Phys. Rev. Lett.}\ }\textbf {\bibinfo {volume} {73}},\ \bibinfo
  {pages} {1376} (\bibinfo {year} {1994})}\BibitemShut {NoStop}%
\bibitem [{\citenamefont {van Megen}\ and\ \citenamefont
  {Underwood}(1993)}]{VanMegen1993}%
  \BibitemOpen
  \bibfield  {author} {\bibinfo {author} {\bibfnamefont {W.}~\bibnamefont {van
  Megen}}\ and\ \bibinfo {author} {\bibfnamefont {S.~M.}\ \bibnamefont
  {Underwood}},\ }\bibfield  {title} {\bibinfo {title} {Glass transition in
  colloidal hard spheres: Mode-coupling theory analysis},\ }\href
  {https://doi.org/10.1103/PhysRevLett.70.2766} {\bibfield  {journal} {\bibinfo
   {journal} {Phys. Rev. Lett.}\ }\textbf {\bibinfo {volume} {70}},\ \bibinfo
  {pages} {2766} (\bibinfo {year} {1993})}\BibitemShut {NoStop}%
\bibitem [{\citenamefont {Fuchs}\ \emph {et~al.}(1998)\citenamefont {Fuchs},
  \citenamefont {G\"otze},\ and\ \citenamefont {Mayr}}]{Fuchs1998}%
  \BibitemOpen
  \bibfield  {author} {\bibinfo {author} {\bibfnamefont {M.}~\bibnamefont
  {Fuchs}}, \bibinfo {author} {\bibfnamefont {W.}~\bibnamefont {G\"otze}},\
  and\ \bibinfo {author} {\bibfnamefont {M.~R.}\ \bibnamefont {Mayr}},\
  }\bibfield  {title} {\bibinfo {title} {Asymptotic laws for tagged-particle
  motion in glassy systems},\ }\href {https://doi.org/10.1103/PhysRevE.58.3384}
  {\bibfield  {journal} {\bibinfo  {journal} {Phys. Rev. E}\ }\textbf {\bibinfo
  {volume} {58}},\ \bibinfo {pages} {3384} (\bibinfo {year}
  {1998})}\BibitemShut {NoStop}%
\bibitem [{\citenamefont {Jung}\ \emph
  {et~al.}(2020{\natexlab{c}})\citenamefont {Jung}, \citenamefont {Voigtmann},\
  and\ \citenamefont {Franosch}}]{Jung:2020_scalingtheory}%
  \BibitemOpen
  \bibfield  {author} {\bibinfo {author} {\bibfnamefont {G.}~\bibnamefont
  {Jung}}, \bibinfo {author} {\bibfnamefont {T.}~\bibnamefont {Voigtmann}},\
  and\ \bibinfo {author} {\bibfnamefont {T.}~\bibnamefont {Franosch}},\
  }\bibfield  {title} {\bibinfo {title} {Scaling equations for mode-coupling
  theories with multiple decay channels},\ }\href
  {https://doi.org/10.1088/1742-5468/ab9e61} {\bibfield  {journal} {\bibinfo
  {journal} {Journal of Statistical Mechanics: Theory and Experiment}\ }\textbf
  {\bibinfo {volume} {2020}},\ \bibinfo {pages} {073301} (\bibinfo {year}
  {2020}{\natexlab{c}})}\BibitemShut {NoStop}%
\bibitem [{\citenamefont {Phillips}(1996)}]{phillips1996stretched}%
  \BibitemOpen
  \bibfield  {author} {\bibinfo {author} {\bibfnamefont {J.}~\bibnamefont
  {Phillips}},\ }\bibfield  {title} {\bibinfo {title} {Stretched exponential
  relaxation in molecular and electronic glasses},\ }\href@noop {} {\bibfield
  {journal} {\bibinfo  {journal} {Reports on Progress in Physics}\ }\textbf
  {\bibinfo {volume} {59}},\ \bibinfo {pages} {1133} (\bibinfo {year}
  {1996})}\BibitemShut {NoStop}%
\bibitem [{\citenamefont {Franosch}\ \emph
  {et~al.}(1997{\natexlab{a}})\citenamefont {Franosch}, \citenamefont {Fuchs},
  \citenamefont {G\"otze}, \citenamefont {Mayr},\ and\ \citenamefont
  {Singh}}]{Franosch1997}%
  \BibitemOpen
  \bibfield  {author} {\bibinfo {author} {\bibfnamefont {T.}~\bibnamefont
  {Franosch}}, \bibinfo {author} {\bibfnamefont {M.}~\bibnamefont {Fuchs}},
  \bibinfo {author} {\bibfnamefont {W.}~\bibnamefont {G\"otze}}, \bibinfo
  {author} {\bibfnamefont {M.~R.}\ \bibnamefont {Mayr}},\ and\ \bibinfo
  {author} {\bibfnamefont {A.~P.}\ \bibnamefont {Singh}},\ }\bibfield  {title}
  {\bibinfo {title} {Asymptotic laws and preasymptotic correction formulas for
  the relaxation near glass-transition singularities},\ }\href
  {https://doi.org/10.1103/PhysRevE.55.7153} {\bibfield  {journal} {\bibinfo
  {journal} {Phys. Rev. E}\ }\textbf {\bibinfo {volume} {55}},\ \bibinfo
  {pages} {7153} (\bibinfo {year} {1997}{\natexlab{a}})}\BibitemShut {NoStop}%
\bibitem [{\citenamefont {Petzold}\ \emph {et~al.}(2013)\citenamefont
  {Petzold}, \citenamefont {Schmidtke}, \citenamefont {Kahlau}, \citenamefont
  {Bock}, \citenamefont {Meier}, \citenamefont {Micko}, \citenamefont {Kruk},\
  and\ \citenamefont {Rössler}}]{doi:10.1063/1.4770055}%
  \BibitemOpen
  \bibfield  {author} {\bibinfo {author} {\bibfnamefont {N.}~\bibnamefont
  {Petzold}}, \bibinfo {author} {\bibfnamefont {B.}~\bibnamefont {Schmidtke}},
  \bibinfo {author} {\bibfnamefont {R.}~\bibnamefont {Kahlau}}, \bibinfo
  {author} {\bibfnamefont {D.}~\bibnamefont {Bock}}, \bibinfo {author}
  {\bibfnamefont {R.}~\bibnamefont {Meier}}, \bibinfo {author} {\bibfnamefont
  {B.}~\bibnamefont {Micko}}, \bibinfo {author} {\bibfnamefont
  {D.}~\bibnamefont {Kruk}},\ and\ \bibinfo {author} {\bibfnamefont {E.~A.}\
  \bibnamefont {Rössler}},\ }\bibfield  {title} {\bibinfo {title} {Evolution
  of the dynamic susceptibility in molecular glass formers: Results from light
  scattering, dielectric spectroscopy, and nmr},\ }\href
  {https://doi.org/10.1063/1.4770055} {\bibfield  {journal} {\bibinfo
  {journal} {The Journal of Chemical Physics}\ }\textbf {\bibinfo {volume}
  {138}},\ \bibinfo {pages} {12A510} (\bibinfo {year} {2013})}\BibitemShut
  {NoStop}%
\bibitem [{\citenamefont {G{\"{o}}tze}(1990)}]{Gotze_1990}%
  \BibitemOpen
  \bibfield  {author} {\bibinfo {author} {\bibfnamefont {W.}~\bibnamefont
  {G{\"{o}}tze}},\ }\bibfield  {title} {\bibinfo {title} {{The scaling
  functions for the $\beta$-relaxation process of supercooled liquids and
  glasses}},\ }\href {https://doi.org/10.1088/0953-8984/2/42/025} {\bibfield
  {journal} {\bibinfo  {journal} {Journal of Physics: Condensed Matter}\
  }\textbf {\bibinfo {volume} {2}},\ \bibinfo {pages} {8485} (\bibinfo {year}
  {1990})}\BibitemShut {NoStop}%
\bibitem [{\citenamefont {Kob}\ and\ \citenamefont
  {Andersen}(1995)}]{PhysRevE.52.4134}%
  \BibitemOpen
  \bibfield  {author} {\bibinfo {author} {\bibfnamefont {W.}~\bibnamefont
  {Kob}}\ and\ \bibinfo {author} {\bibfnamefont {H.~C.}\ \bibnamefont
  {Andersen}},\ }\bibfield  {title} {\bibinfo {title} {Testing mode-coupling
  theory for a supercooled binary {Lennard-Jones} mixture. ii. intermediate
  scattering function and dynamic susceptibility},\ }\href
  {https://doi.org/10.1103/PhysRevE.52.4134} {\bibfield  {journal} {\bibinfo
  {journal} {Phys. Rev. E}\ }\textbf {\bibinfo {volume} {52}},\ \bibinfo
  {pages} {4134} (\bibinfo {year} {1995})}\BibitemShut {NoStop}%
\bibitem [{\citenamefont {Fuchs}\ \emph {et~al.}(1992)\citenamefont {Fuchs},
  \citenamefont {Hofacker},\ and\ \citenamefont {Latz}}]{PhysRevA.45.898}%
  \BibitemOpen
  \bibfield  {author} {\bibinfo {author} {\bibfnamefont {M.}~\bibnamefont
  {Fuchs}}, \bibinfo {author} {\bibfnamefont {I.}~\bibnamefont {Hofacker}},\
  and\ \bibinfo {author} {\bibfnamefont {A.}~\bibnamefont {Latz}},\ }\bibfield
  {title} {\bibinfo {title} {Primary relaxation in a hard-sphere system},\
  }\href {https://doi.org/10.1103/PhysRevA.45.898} {\bibfield  {journal}
  {\bibinfo  {journal} {Phys. Rev. A}\ }\textbf {\bibinfo {volume} {45}},\
  \bibinfo {pages} {898} (\bibinfo {year} {1992})}\BibitemShut {NoStop}%
\bibitem [{\citenamefont {Fuchs}(1994)}]{FUCHS1994241}%
  \BibitemOpen
  \bibfield  {author} {\bibinfo {author} {\bibfnamefont {M.}~\bibnamefont
  {Fuchs}},\ }\bibfield  {title} {\bibinfo {title} {The kohlrausch law as a
  limit solution to mode coupling equations},\ }\href
  {https://doi.org/https://doi.org/10.1016/0022-3093(94)90442-1} {\bibfield
  {journal} {\bibinfo  {journal} {Journal of Non-Crystalline Solids}\ }\textbf
  {\bibinfo {volume} {172-174}},\ \bibinfo {pages} {241} (\bibinfo {year}
  {1994})},\ \bibinfo {note} {relaxations in Complex Systems}\BibitemShut
  {NoStop}%
\bibitem [{\citenamefont {Fuchs}(1995)}]{doi:10.1080/00411459508203937}%
  \BibitemOpen
  \bibfield  {author} {\bibinfo {author} {\bibfnamefont {M.}~\bibnamefont
  {Fuchs}},\ }\bibfield  {title} {\bibinfo {title} {{MCT} results for a simple
  liquid at the glass transition},\ }\href
  {https://doi.org/10.1080/00411459508203937} {\bibfield  {journal} {\bibinfo
  {journal} {Transport Theory and Statistical Physics}\ }\textbf {\bibinfo
  {volume} {24}},\ \bibinfo {pages} {855} (\bibinfo {year} {1995})}\BibitemShut
  {NoStop}%
\bibitem [{\citenamefont {Weysser}\ \emph {et~al.}(2010)\citenamefont
  {Weysser}, \citenamefont {Puertas}, \citenamefont {Fuchs},\ and\
  \citenamefont {Voigtmann}}]{PhysRevE.82.011504}%
  \BibitemOpen
  \bibfield  {author} {\bibinfo {author} {\bibfnamefont {F.}~\bibnamefont
  {Weysser}}, \bibinfo {author} {\bibfnamefont {A.~M.}\ \bibnamefont
  {Puertas}}, \bibinfo {author} {\bibfnamefont {M.}~\bibnamefont {Fuchs}},\
  and\ \bibinfo {author} {\bibfnamefont {T.}~\bibnamefont {Voigtmann}},\
  }\bibfield  {title} {\bibinfo {title} {Structural relaxation of polydisperse
  hard spheres: Comparison of the mode-coupling theory to a langevin dynamics
  simulation},\ }\href {https://doi.org/10.1103/PhysRevE.82.011504} {\bibfield
  {journal} {\bibinfo  {journal} {Phys. Rev. E}\ }\textbf {\bibinfo {volume}
  {82}},\ \bibinfo {pages} {011504} (\bibinfo {year} {2010})}\BibitemShut
  {NoStop}%
\bibitem [{\citenamefont {Gruber}(2019)}]{Gruber2019}%
  \BibitemOpen
  \bibfield  {author} {\bibinfo {author} {\bibfnamefont {M.}~\bibnamefont
  {Gruber}},\ }\emph {\bibinfo {title} {Theory of microrheology in complex
  fluids}},\ \href {http://nbn-resolving.de/urn:nbn:de:bsz:352-2-hurftyctohii2}
  {Ph.D. thesis},\ \bibinfo  {school} {Universit{\"a}t Konstanz}, \bibinfo
  {address} {Konstanz} (\bibinfo {year} {2019})\BibitemShut {NoStop}%
\bibitem [{\citenamefont {Gruber}\ \emph {et~al.}(2020)\citenamefont {Gruber},
  \citenamefont {Puertas},\ and\ \citenamefont {Fuchs}}]{PhysRevE.101.012612}%
  \BibitemOpen
  \bibfield  {author} {\bibinfo {author} {\bibfnamefont {M.}~\bibnamefont
  {Gruber}}, \bibinfo {author} {\bibfnamefont {A.~M.}\ \bibnamefont
  {Puertas}},\ and\ \bibinfo {author} {\bibfnamefont {M.}~\bibnamefont
  {Fuchs}},\ }\bibfield  {title} {\bibinfo {title} {Critical force in active
  microrheology},\ }\href {https://doi.org/10.1103/PhysRevE.101.012612}
  {\bibfield  {journal} {\bibinfo  {journal} {Phys. Rev. E}\ }\textbf {\bibinfo
  {volume} {101}},\ \bibinfo {pages} {012612} (\bibinfo {year}
  {2020})}\BibitemShut {NoStop}%
\bibitem [{\citenamefont {Liluashvili}\ \emph {et~al.}(2017)\citenamefont
  {Liluashvili}, \citenamefont {{\'O}nody},\ and\ \citenamefont
  {Voigtmann}}]{liluashvili2017mode}%
  \BibitemOpen
  \bibfield  {author} {\bibinfo {author} {\bibfnamefont {A.}~\bibnamefont
  {Liluashvili}}, \bibinfo {author} {\bibfnamefont {J.}~\bibnamefont
  {{\'O}nody}},\ and\ \bibinfo {author} {\bibfnamefont {T.}~\bibnamefont
  {Voigtmann}},\ }\bibfield  {title} {\bibinfo {title} {Mode-coupling theory
  for active brownian particles},\ }\href@noop {} {\bibfield  {journal}
  {\bibinfo  {journal} {Physical Review E}\ }\textbf {\bibinfo {volume} {96}},\
  \bibinfo {pages} {062608} (\bibinfo {year} {2017})}\BibitemShut {NoStop}%
\bibitem [{\citenamefont {Reichert}\ \emph {et~al.}(2021)\citenamefont
  {Reichert}, \citenamefont {Granz},\ and\ \citenamefont
  {Voigtmann}}]{reichert2021transport}%
  \BibitemOpen
  \bibfield  {author} {\bibinfo {author} {\bibfnamefont {J.}~\bibnamefont
  {Reichert}}, \bibinfo {author} {\bibfnamefont {L.~F.}\ \bibnamefont
  {Granz}},\ and\ \bibinfo {author} {\bibfnamefont {T.}~\bibnamefont
  {Voigtmann}},\ }\bibfield  {title} {\bibinfo {title} {Transport coefficients
  in dense active brownian particle systems: mode-coupling theory and
  simulation results},\ }\href@noop {} {\bibfield  {journal} {\bibinfo
  {journal} {The European Physical Journal E}\ }\textbf {\bibinfo {volume}
  {44}},\ \bibinfo {pages} {1} (\bibinfo {year} {2021})}\BibitemShut {NoStop}%
\bibitem [{\citenamefont {Schilling}\ and\ \citenamefont
  {Scheidsteger}(1997)}]{schilling1997mode}%
  \BibitemOpen
  \bibfield  {author} {\bibinfo {author} {\bibfnamefont {R.}~\bibnamefont
  {Schilling}}\ and\ \bibinfo {author} {\bibfnamefont {T.}~\bibnamefont
  {Scheidsteger}},\ }\bibfield  {title} {\bibinfo {title} {Mode coupling
  approach to the ideal glass transition of molecular liquids: Linear
  molecules},\ }\href@noop {} {\bibfield  {journal} {\bibinfo  {journal}
  {Physical Review E}\ }\textbf {\bibinfo {volume} {56}},\ \bibinfo {pages}
  {2932} (\bibinfo {year} {1997})}\BibitemShut {NoStop}%
\bibitem [{\citenamefont {Franosch}\ \emph
  {et~al.}(1997{\natexlab{b}})\citenamefont {Franosch}, \citenamefont {Fuchs},
  \citenamefont {G\"otze}, \citenamefont {Mayr},\ and\ \citenamefont
  {Singh}}]{PhysRevE.56.5659}%
  \BibitemOpen
  \bibfield  {author} {\bibinfo {author} {\bibfnamefont {T.}~\bibnamefont
  {Franosch}}, \bibinfo {author} {\bibfnamefont {M.}~\bibnamefont {Fuchs}},
  \bibinfo {author} {\bibfnamefont {W.}~\bibnamefont {G\"otze}}, \bibinfo
  {author} {\bibfnamefont {M.~R.}\ \bibnamefont {Mayr}},\ and\ \bibinfo
  {author} {\bibfnamefont {A.~P.}\ \bibnamefont {Singh}},\ }\bibfield  {title}
  {\bibinfo {title} {Theory for the reorientational dynamics in glass-forming
  liquids},\ }\href {https://doi.org/10.1103/PhysRevE.56.5659} {\bibfield
  {journal} {\bibinfo  {journal} {Phys. Rev. E}\ }\textbf {\bibinfo {volume}
  {56}},\ \bibinfo {pages} {5659} (\bibinfo {year}
  {1997}{\natexlab{b}})}\BibitemShut {NoStop}%
\bibitem [{\citenamefont {G\"otze}\ \emph {et~al.}(2000)\citenamefont
  {G\"otze}, \citenamefont {Singh},\ and\ \citenamefont
  {Voigtmann}}]{PhysRevE.61.6934}%
  \BibitemOpen
  \bibfield  {author} {\bibinfo {author} {\bibfnamefont {W.}~\bibnamefont
  {G\"otze}}, \bibinfo {author} {\bibfnamefont {A.~P.}\ \bibnamefont {Singh}},\
  and\ \bibinfo {author} {\bibfnamefont {T.}~\bibnamefont {Voigtmann}},\
  }\bibfield  {title} {\bibinfo {title} {Reorientational relaxation of a linear
  probe molecule in a simple glassy liquid},\ }\href
  {https://doi.org/10.1103/PhysRevE.61.6934} {\bibfield  {journal} {\bibinfo
  {journal} {Phys. Rev. E}\ }\textbf {\bibinfo {volume} {61}},\ \bibinfo
  {pages} {6934} (\bibinfo {year} {2000})}\BibitemShut {NoStop}%
\bibitem [{\citenamefont {Sperl}(2000)}]{Sperl2000}%
  \BibitemOpen
  \bibfield  {author} {\bibinfo {author} {\bibfnamefont {M.}~\bibnamefont
  {Sperl}},\ }\href@noop {} {\emph {\bibinfo {title} {{Glass transition in
  colloids with attractive interaction. Diploma thesis, Technische
  Universit{\"{a}}t M{\"{u}}nchen}}}}\ (\bibinfo {year} {2000})\BibitemShut
  {NoStop}%
\end{thebibliography}%

\end{document}